\documentclass[showpacs,twocolumn,prd,nofootinbib,reprint]{revtex4-2}
\usepackage{adjustbox}
\usepackage[utf8]{inputenc}
 \usepackage{color,graphicx,epsfig}
 \usepackage{amsmath,amssymb,dsfont,epsfig,graphicx,xcolor,fontenc}
 \usepackage{ifpdf}
 \usepackage{amsmath}
 \usepackage{bm}
 \usepackage[english]{babel}
 \usepackage{graphicx}%
 \usepackage{amsfonts}%
 \usepackage{amssymb}
 \usepackage{braket}
 \usepackage{hyperref}
 \usepackage{epstopdf}
 \usepackage{ulem}
\usepackage{appendix}

\definecolor{nicered}{rgb}{0.7,0.1,0.1}
\definecolor{nicegreen}{rgb}{0.1,0.5,0.1}
\definecolor{violet}{rgb}{0.7,0.3,0.3}
\hypersetup{colorlinks,citecolor=nicegreen,linkcolor=nicered}

\def\mysection#1{{{\vspace{2mm}\noindent\bf #1}~}}
\def\LjubljanaFMF{Faculty of Mathematics and Physics, University of Ljubljana,
 Jadranska 19, 1000 Ljubljana, Slovenia }
\def\LjubljanaIJS{Jo\v zef Stefan Institute, Jamova 39, 1000 Ljubljana, Slovenia}
\def\Heidelberg{Institut f{\"u}r Theoretische Physik, Universit{\"a}t Heidelberg, Germany}
\def\Argentina{International Center for Advanced Studies (ICAS) and CONICET, UNSAM,
Campus Miguelete, 25 de Mayo y Francia, CP1650, San Martin, Buenos Aires, Argentina}
\def\Argentinab{Centro At\'omico Bariloche, Instituto Balseiro and CONICET}
\def\Zurich{Physik-Institut, Universit\"at Z\"urich, CH-8057, Switzerland}

\begin{document}

\title{Bayesian Probabilistic Modelling for Four-Tops at the LHC}

\author{Ezequiel Alvarez}
\email[Electronic address:]{sequi@unsam.edu.ar}
\affiliation{\Argentina}

\author{Barry~M.~Dillon}
\email[Electronic address:]{dillon@thphys.uni-heidelberg.de} 
\affiliation{\Heidelberg}

\author{Darius~A.~Faroughy}
\email[Electronic address:]{faroughy@physik.uzh.ch}
\affiliation{\Zurich}

\author{Jernej~F.~Kamenik}
\email[Electronic address:]{jernej.kamenik@cern.ch} 
\affiliation{\LjubljanaIJS}
\affiliation{\LjubljanaFMF}

\author{Federico Lamagna}
\email[Electronic address:]{federico.lamagna@cab.cnea.gov.ar}
\affiliation{\Argentinab}

\author{Manuel~Szewc}
\email[Electronic address:]{mszewc@unsam.edu.ar}
\affiliation{\Argentina}
\affiliation{\LjubljanaIJS}

\begin{abstract}


Monte Carlo (MC) generators are crucial for analyzing data in particle collider experiments. However, often even a small mismatch between the MC simulations and the measurements can undermine the interpretation of the results. This is particularly important in the context of LHC searches for rare physics processes within and beyond the standard model (SM). 
One of the ultimate rare processes in the SM currently being explored at the LHC, $pp\to t\bar tt \bar t$ with its large multi-dimensional phase-space is an ideal testing ground to explore new ways to reduce the impact of potential MC mismodelling on experimental results.
We propose a novel statistical method capable of disentangling the 4-top signal from the dominant backgrounds in the same-sign dilepton channel, while simultaneously correcting for possible MC imperfections in modelling of the most relevant discriminating observables -- the jet multiplicity distributions. A Bayesian mixture of multinomials is used to model the light-jet and $b$-jet multiplicities under the assumption of their conditional independence.
The signal and background distributions generated from a deliberately mistuned MC simulator are used as model priors. The posterior distributions, as well as the signal and background fractions, are then learned from the data using Bayesian inference. We demonstrate  that our method can mitigate the effects of large MC mismodellings in the context of a realistic $t\bar tt\bar t$ search, leading to corrected posterior distributions that better approximate the underlying truth-level spectra.

\end{abstract}

\maketitle

\section{Introduction}
In recent years, the large abundance of LHC data on one hand, and the absence of clear New Physics (NP) signals in theory driven analyses of this data on the other, have motivated the development of novel, more data driven approaches to LHC data analysis and NP searches. In particular, the advent of unsupervised and weakly-supervised Machine Learning (ML) techniques has allowed for the development of broad model independent NP search and characterisation strategies~\cite{Kasieczka:2021xcg}. Simultaneously, there have been important efforts to reduce reliance of LHC measurements on Monte Carlo (MC) simulations of hadronic processes~\cite{Kasieczka:2020pil,Ghosh:2021roe,Benkendorfer:2020gek,Choi:2020bnf, Flesher:2020kuy}. 

The simultaneous production of four top quarks represents an important NP benchmark (see e.g. Refs.~\cite{Lillie:2007hd,Kumar:2009vs,Acharya:2009gb,Kim:2016plm,Liu:2015hxi,AguilarSaavedra:2011ck, Camargo-Molina:2018cwu, Alvarez:2019uxp, Darme:2021xxu,Khatibi:2020mvt,Banelli:2020iau,Cao:2021qqt}), but also an interesting point of coalescence for several of these developments~\cite{Alvarez:2019knh}. One of the main issues in studying this final state is its tiny cross-section (12 fb) compared to its main backgrounds ($\sim600$ fb), which is compounded by the challenges to correctly model the complex final states through MC simulations. To address these issues, we have previously studied the two lepton same sign channel (2LSS${\pm\pm}$)~\cite{Alvarez:2016nrz} which in the SM may contain signal and background events up to the same order of magnitude and furthermore exhibits somewhat reduced complexity of the (multi jet) final state, compared to the single lepton channel~\cite{ATLAS:2021kqb,CMS:2019jsc}. In the 2LSS++ channel $t \bar t W^+$ production represents the main and most challenging background for the 4-top signal.\footnote{Our results and discussion would apply equally well to other non-negligible backgrounds such as $t\bar t h$ and $t\bar t Z$.} Recent experimental analyses in this channel~\cite{Sirunyan:2019wxt, Aad:2020klt} have highlighted difficulties in reliably modelling the signal and background kinematics using state of the art MC tools. This in turn hinders the sensitivity of this important signature to possible NP effects in four-top production.

Using the experimental challenge described above as an example and motivation, in the present paper we describe a novel Bayesian statistical framework to disentangle in-situ signal and background distributions of categorical data. Our method can be used to simultaneously identify and correct potential (MC) mismodelling of discrete distributions as well as extract signal and background admixtures in the data close to their truth values.  

The paper is organized as follows. In Sec.~\ref{sec:Mixture} we introduce our statistical model of multinomial mixtures with Bayesian inference and demonstrate its use on a toy example. We apply the model to jet multiplicity distributions in the 2LSS++ channel of 4-top production at the LHC in Sec.~\ref{sec:4t} and show how it can be used to identify and correct MC mismodelling and extract signal and background fractions. Sec.~\ref{sec:conditional} is devoted to a detailed study of the assumptions and consistency checks of the model when applied to realistic datasets. Finally, we summarize our findings in Sec.~\ref{sec:Conclusions}.

\section{Categorical mixture model for four-tops}\label{sec:Mixture}

Anticipating the application to 4-top production, in the following we represent an event generation process by a pair of random variables $(N_j,N_b)$ indicating the number of clustered light-jets and $b$-jets, respectively. 
Our starting point is that a collection of such events can be described using a likelihood with a joint probability density $p(j,b)$ where $j$ ($b$) are the observed number of light-jets (b-jets) in an event. The most general discrete model for this likelihood is the multinomial distribution\footnote{Along this work we refer to multinomial distribution although in all cases it consists of a single drawing per event and therefore it is also a categorical distribution, which is a special case of the former.} with $d_{j}\times d_{b}-1$ parameters, where $d_{j,b}$ are the number of possible light-jets and $b$-jets to be expected in an event. 
However, our goal is to disentangle the contributions to this joint likelihood arising from four-top events and $t\bar{t}W$ events.
To do so we introduce two mixture components, one for $t\bar{t}W$ and one for four-top.
If we simply describe each mixture with a multinomial distribution $p(j,b|z)$ with $z\in[0,1]$ representing the mixture label, we would have a mixture model with $2\times(d_j\times d_b-1)+1$ parameters.
Since each event is independent and consists of just a single draw from this distribution, each mixture can describe all possible combinations of $N_j$ and $N_b$ values in the data and therefore all correlations by itself.
The model would thus over-parameterize the data making the inclusion of mixtures redundant.\footnote{Note that this would not be the case if each event was generated by several draws from $p(j,b|z)$, since there would then be additional correlations between the multiple draws per event. This is the case in s.c. mixed membership models~\cite{Dillon:2019cqt,Dillon:2020quc,Dillon:2021nxw} used in jet substructure analyses where the mixtures describe correlations between the multiple draws per event.}

Therefore the key insight is to instead write down a mixture model in terms of $p(j|z)$ and $p(b|z)$, such that the correlations between $N_{j}$ and $N_{b}$ in the dataset are parameterized by the class label alone. The number of parameters in this model is $2\times (d_j+d_b-2)+1$. To be explicit, we optimize the model to parameterize the correlations between $N_j$ and $N_b$ in terms of a discrete variable $Z$, and interpret this as a class label for four-top and $t\bar{t}W$ events. We are making the simplifying assumption that $N_j$ and $N_b$ are {\it conditionally independent variables}, that all correlations between them in the dataset arise only from assignments to the two classes. Conditional independence is of course an approximation. In particular, in a realistic measurement setting, $N_j$ and
$N_b$ are not strictly conditionally independent due to mis-tagging or other
reconstruction imperfections.  
The degree to which the method succeeds is limited by this approximation. Conversely, a failure of the method to converge to a consistent description of the measured distributions would be a clear sign that the assumptions of the statistical model are not respected by the dataset. We return to this important caveat and discuss its mitigation in Sec.~\ref{sec:conditional}.\footnote{A systematic study of statistical models which go beyond strict conditional independence assumptions is in progress and will be presented elsewhere.} However, as we will show, in the case at hand, the method exhibits good convergence indicating that conditional independence holds sufficiently well in practice. \\

Within the limitations described above, the generative process for the dataset proceeds as follows: for each event ($n$) a class label $z_n$ is first drawn from a binomial probability distribution parametrized by $\pi\in[0,1]$. Then $j_n$ and $b_n$ are sampled from separate multinomials corresponding to the drawn class and parametrized by $\alpha_{z,i}$ and $\beta_{z,k}$, respectively, where $i$ and $k$ run up to $d_j$ and $d_b$, respectively.  We assume that the whole dataset $X$, consisting of $n \in N$ pairs of measurements $x_{n}$ = ($j_n$, $b_n$) for the 2LSS++ selected events, is generated through this probabilistic model and we want to infer the values of its parameters, namely $\pi, \alpha_{0,j}, \beta_{0,i},\alpha_{1,j}$ and $\beta_{1,i}$, which we collectively indicate as $\theta$.  Observe that the described model corresponds to a special case of a {\it mixture of multinomials}~\cite{Bishop:998831}. 

\begin{figure}
    \centering
    \includegraphics[width=0.5\textwidth]{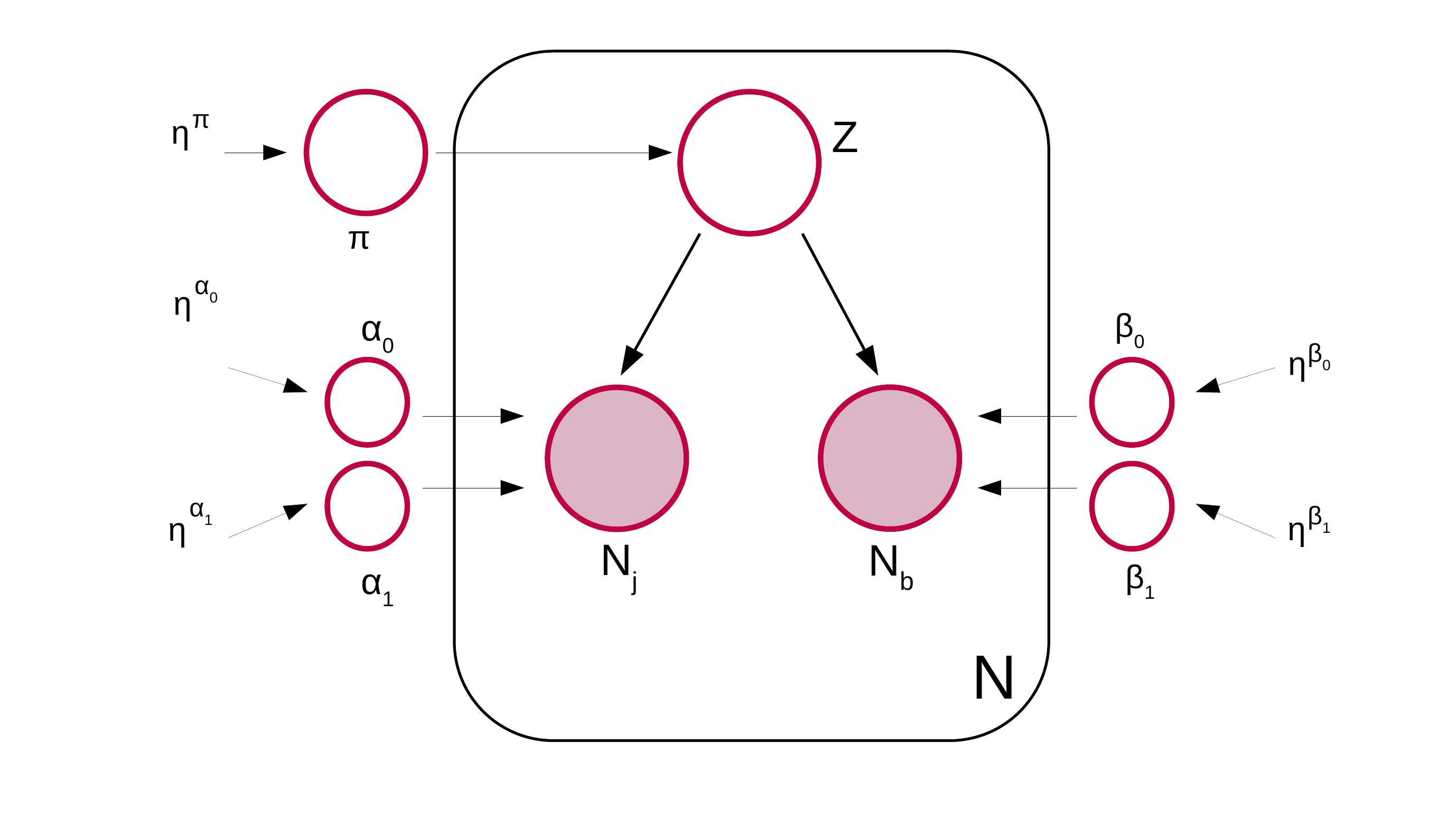}
    \caption{Plate diagram of (Bayesian) 2-mixture model of multinomials for $(N_b,N_j)$ $N$-event dataset. From the Dirichlet prior distributions (with hyperparameters $\eta_i$) the multinomial parameters ($\pi,\ \alpha_i$ and $\beta_i$) are sampled, then $N$ events are sampled through a latent variable $Z$ that determines in turn from which multinomial the two observables in each event ($N_j$ and $N_b$) are sampled.
    }
    \label{graphical_model}
\end{figure}

Adopting a Bayesian framework, we consider the model parameters ($\theta$) to be random variables as well and we want to update our knowledge of these random variables after measuring $X$. However, it is more convenient in practice to consider explicitly also the latent variables $Z$ which represent the class assignments of each event. Graphically, the probabilistic model can be represented through the plate diagram in Fig.~\ref{graphical_model} and leads to the posterior:
\begin{equation}
    p(Z,\pi,\alpha,\beta|X)=\frac{p(X,Z,\pi,\alpha,\beta)}{p(X)}\,,
    \label{posterior_bis}
\end{equation}
where the joint distribution $p(X,Z,\pi,\alpha,\beta)$ is given explicitly by
\begin{eqnarray}
    p(X,Z,\pi,\alpha,\beta)&=&\prod_{n=1}^{N}p(x_{n}|z_{n},\alpha,\beta)p(z_{n}|\pi)\nonumber
    \\&&p(\pi|\eta_{\pi})\prod_{k=0}^{1}p(\alpha_{k}|\eta_{\alpha_{k}})p(\beta_{k}|\eta_{\beta_{k}})\,.\nonumber
    \label{joint_distribution}
\end{eqnarray}
Here $p(x_{n}|z_{n},\alpha,\beta)=\alpha_{z_{n}j_{n}}\beta_{z_{n}b_{n}}$, $p(z_{n}|\pi)=\pi_{z_{n}}$ and $p(\pi|\eta^{\pi})$, $p(\alpha_{k}|\eta^{\alpha_{k}})$ and $p(\beta_{k}|\eta^{\beta_{k}})$ are Dirichlet distributions with the corresponding $\eta^i$ set of parameters.

The main idea in this expression is that given the dataset $X$, a probabilistic model that allows us to write down an expression for $p(X|\theta)$ and a reasonable prior $p(\theta)$, we can in principle determine the probability density function (pdf) for the parameters $p(\theta|X)$.  This is a powerful result, since it gives us not only the fraction of signal to background and its uncertainty through $p(\pi|X)$ marginalizing over the other parameters, but it can also give us the $N_j$ and $N_b$ distributions of both individual classes.  If the probabilistic model describes  the data well and the prior is reasonable, then these should match within uncertainties the true underlying background and signal $N_j$ and $N_b$ distributions. 

There are many known approaches to solving Eq.~\eqref{posterior_bis} using Bayesian Inference; including mean-field techniques such as Variational Inference (VI)~\cite{Bishop:998831} and numerical Markov Chain Monte Carlo methods such as Gibbs Sampling (GS)~\cite{Bishop:998831}. Below we focus on the latter numerical approach which turns out to be preferred to the mean-field methods which approximate the posterior with a fully factorized model that neglects possible correlations between the inferred parameters. As we are interested in finding the correlations between $N_{j}$ and $N_{b}$ through class assignment, VI is challenged by definition to find the appropriate correlations.

The goal of the GS algorithm is to approximate the posterior through the use of a finite number of samples. These samples can then be used to obtain any desired expected values such as the mean of the relevant parameters $\mathbb{E}[\theta_{i}]$. To obtain samples from the posterior, each iteration samples an observation of each parameter $\theta_{i}$ from the marginal distribution conditioned on the remaining parameters $p(\theta_{i}|\theta_{\setminus i},X)$. When implementing a Gibbs sampler to approximate Eq.~\eqref{posterior_bis}, the conditional distributions can be obtained and sampled from efficiently, being either Dirichlet or Multinomial distributions. Our algorithm implemented in python is available at {\tt GitHub}~\cite{code-bayes-four-tops}.

In practice, subsequently drawn samples are highly correlated. To mitigate this we drop the first $M$ samples, which constitute what is called the burn-in phase, and then apply a `thinning' procedure which consists in only keeping every $l^{\text{th}}$ sample. We also implement different chains, or walkers, initialized at different randomly chosen starting points. We estimate sufficient $M$ and $l$ values by computing the integrated autocorrelation time $\tau$ as defined in Ref.~\cite{Sokal1996MonteCM} and adapting its implementation in {\tt emcee}~\cite{Foreman_Mackey_2013} accounting for the fact that we do not have an ensemble sampler. 
We find that with $30$ walkers and $1000$ saved iterations per walker after thinning with $l=100$ we have $\tau$'s in the range $\tau \in [1,2.5]$. We consider a burn-in phase of $M=1000$ after which we save the aforementioned $1000$ samples with thinning.\footnote{The GS algorithm and techniques described above are well known in other disciplines, in particular computer sciences, however they have to our knowledge not been applied before in the context of (high energy) physics.} Once we have an accurate approximation of $p(Z,\pi,\alpha,\beta|X)$, we can marginalize over the class assignments by neglecting the sampled values $Z$. 

\subsection{A simple toy example}

\begin{figure*}[t]
    \centering
    \includegraphics[width=\textwidth]{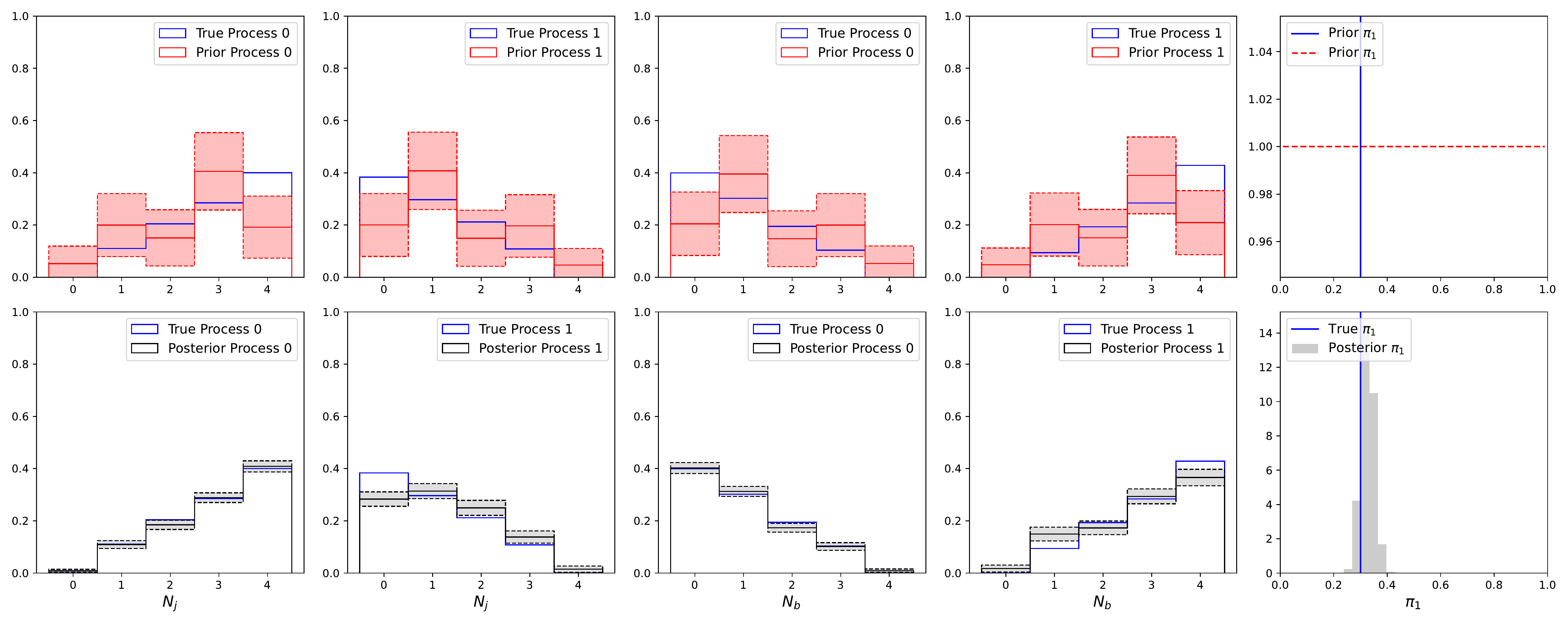}
    \caption{$N_j$, $N_b$ and $\pi_{1}$ distributions: true values (blue), priors (red) and posterior (black) for the toy model.  Shaded regions in first four plots indicate the $1\sigma$ uncertainty region.  Comparing the posteriors to the priors one can appreciate the improvement in estimating the true distributions departing from incorrect and uncertain priors using Bayesian Inference on the data.
    }
    \label{fig:toy_goodprior_dists}
\end{figure*}

To demonstrate the efficiency of this approach, as well as the limitations due to the approximations we make, we will first apply it to inference in a very simple toy model.
We take a sample of `events', each with just two features.
The sample is comprised of two types of events, which for the sake of analogy we call background and signal.
These signal and background events are sampled from sets of overlapping distributions in the feature-space.
The features for each event are sampled independently, therefore in this simple toy example these two features are completely uncorrelated from each other.
We consider the case in which the prior distributions for these features are not too far from the truth. In contrast, we consider a uniform prior distribution for the $\pi$ parameter giving the fraction of signal and background in the sample. This indicates no prior knowledge on how much background and signal we can expect in the dataset and is the most conservative assumption we can make in this regard. We show the prior distributions as well as the true values of the parameters in the upper row of Fig.~\ref{fig:toy_goodprior_dists}

\begin{figure}
    \centering
    \includegraphics[width=0.5\textwidth]{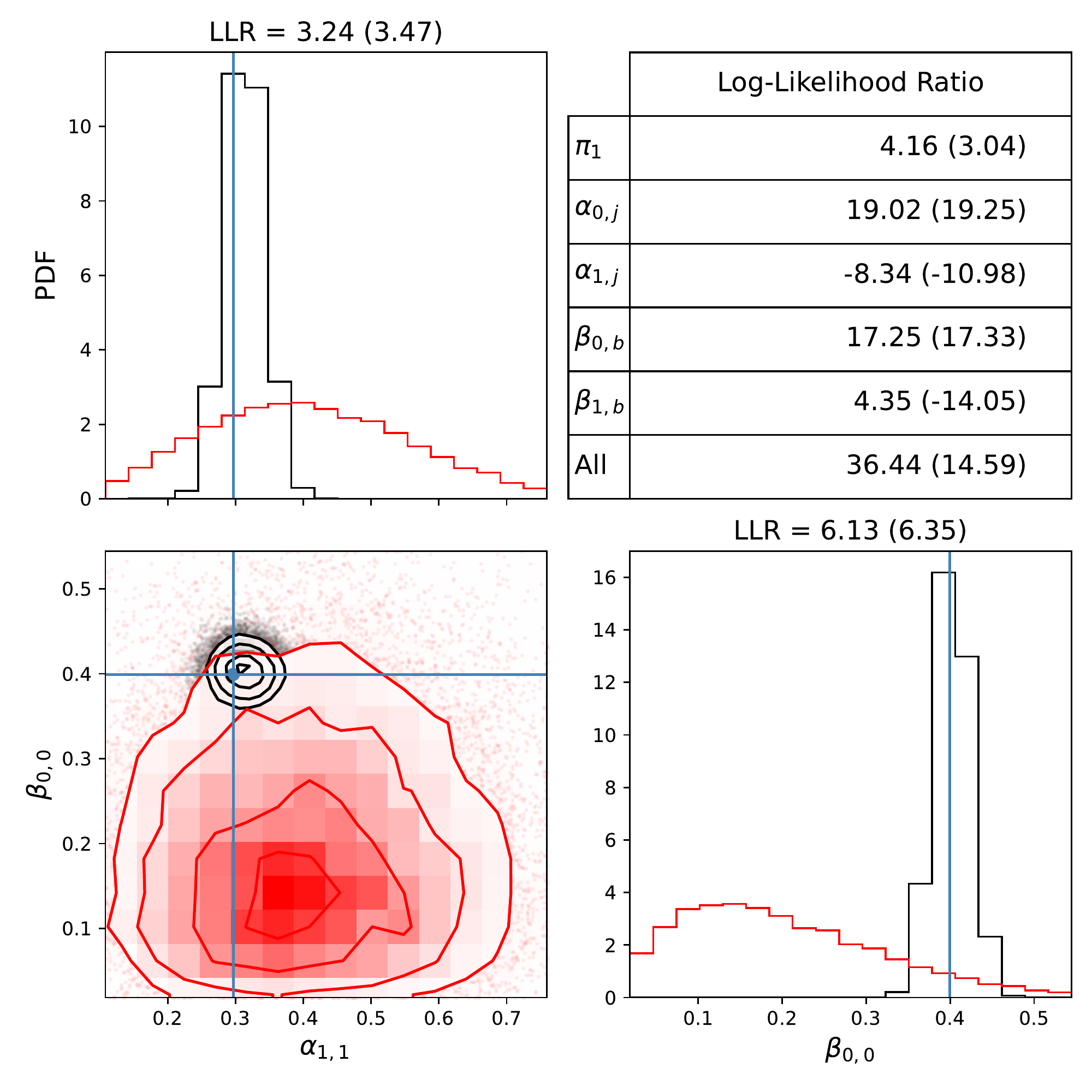}
    \caption{Excerpt from the corner-plot for the toy model.  Red indicates the prior distribution, black the posterior distribution obtained through GS and blue is the true value. We see how the posterior distribution captures the correlation between $N_{j}$ and $N_{b}$. The titles of each 1D histogram contain the Log-Likelihood Ratio between the posterior and the prior using either GS or VI for the posterior estimation, with the latter shown in parentheses. The table contains the sum of Log-Likelihood Ratios per parameter block, again considering the posteriors obtained through GS and through VI. We see that VI is a bad approximation to GS, failing to improve on the prior for several parameter blocks.}
    \label{toy_corner}
\end{figure}

After numerically solving the Bayes Inference problem using GS, we compare the class-$0$ and class-$1$ inferred distributions for $N_j$ and $N_b$ to the truth-level background and signal distributions in $X$. A good summary to assess the success of the algorithm is the corner-plot which visualizes the distribution through marginalizing to either two or one parameter dimensions and the true values. An excerpt is shown in Fig.~\ref{toy_corner}. In each panel we show the corresponding prior distribution (red), posterior distribution (black) and the true values (blue). Quantitatively, one can also compare the level of improvement between the prior and the posterior by computing their Log-Likelihood Ratio (LLR) with respect to the true value for each parameter. We display these numbers above the diagonal panels of the corner-plot, and we see a robust improvement in most of them. To compute the LLR of the posterior and prior of the complete model one would in principle need to evaluate the joint density distributions of all pairs of parameters (off-diagonal elements in the corner-plot) which is beyond the scope of this work. Instead, as a rough approximation, neglecting the correlations between the parameters, we obtain a global LLR as a sum of the individual parameter LLRs, LLR $\approx$ 36. We display this global sum as well as partial sums grouping different parameters together in Fig.~\ref{toy_corner}. We also include the partial and global sums of LLR obtained when approximating the posterior through VI. We observe that although VI captures the maximum of the posterior accurately, it consistently underestimates the variance of the distribution yielding a too narrow approximation to the GS obtained posterior. This is reflected in a lower improvement over the prior (LLR $\approx$ 15). 

Finally, in the bottom row of Fig.~\ref{fig:toy_goodprior_dists} we group together the one-dimensional marginalized posterior distributions for each parameter to obtain the $N_j$, $N_b$ distributions of the signal and background, as well as for the $\pi$ parameter, i.e. the fraction of signal in the sample. In the plot the true value of the parameters is shown in solid blue. Notice that the posterior exhibits good convergence to the true values as well as a considerable reduction of the uncertainty, when compared to the prior, which emulates the imperfect MC. We find an improvement in both the $N_j$ and $N_{b}$ distributions for each process as expected from Fig.~\ref{toy_corner}.  It is also interesting to notice in Fig.~\ref{fig:toy_goodprior_dists} how from a complete ignorance of the signal and background fractions in the sample, the algorithm recovers a pdf for $\pi$ in good agreement with its true value.

\section{Application to four-top measurements}\label{sec:4t}

In the 2LSS++ channel, the final state is usually characterized by at least $2\ell^+$, at least 2 $b$-tagged jets, and at least 4 light jets. Additional cuts on missing transverse energy and transverse momentum may be invoked to enhance the signal fraction in the sample.  The exact details of the event selection are however not important for the purposes of this work. From the decay products at matrix-element level of the signal, one expects {\it a priori} that the $N_j$ and $N_b$ distributions to be skewed towards higher values when compared to the background process, thus providing enough separation for disentangling them using statistical inference.

\begin{figure*}[t]
    \centering
    \includegraphics[width=\textwidth]{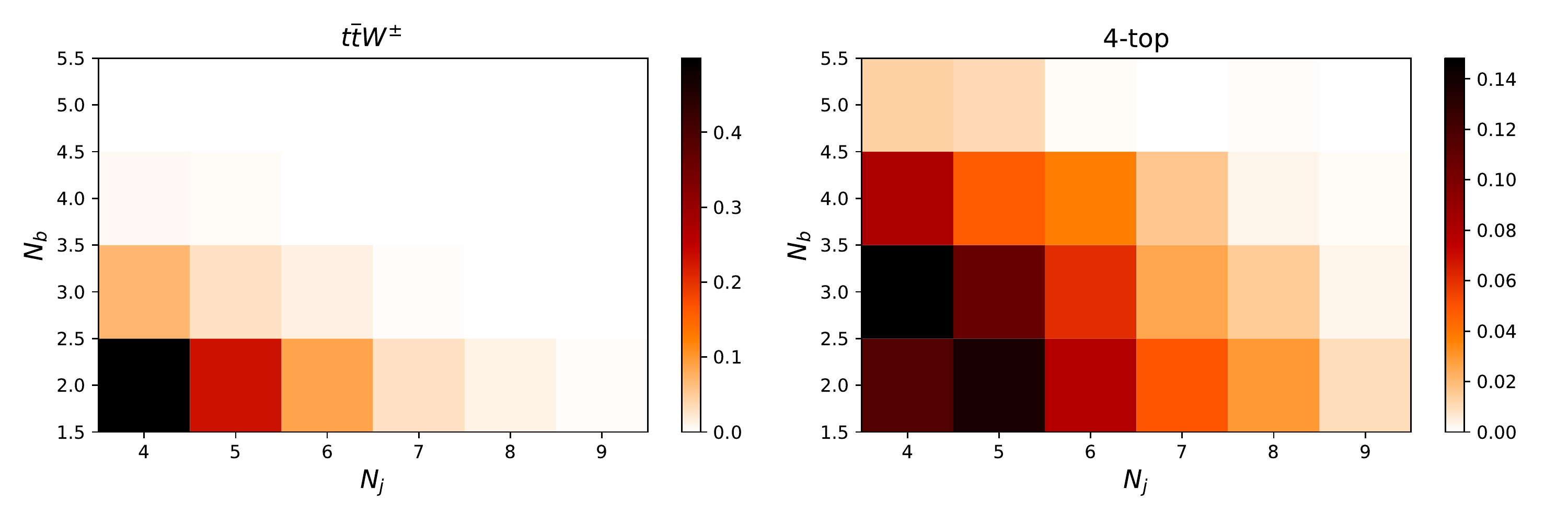}
    \caption{$N_j$ and $N_b$ two dimensional distributions for $t\bar t W^{\pm}$ and 4-top.}
    \label{2d_heat}
\end{figure*}

In our setup we have simulated 4-top and $t\bar t W^\pm$ events using {\tt Madgraph}~\cite{Alwall:2014hca}, {\tt Pythia}~\cite{Sjostrand:2014zea} and {\tt Delphes}~\cite{deFavereau:2013fsa} to account for matrix level calculations and showering, hadronization and detector simulation, respectively. We selected $N=500$ events, roughly equivalent to $\mathcal{L} = 800\mathrm{fb}^{-1}$, in the 2LSS++ channel with 70\% background and 30\% signal (we also tested for other signal fractions and obtained similar results). Using this data we created a dataset $X$, represented by $N$ pairs ($j_n,b_n$), $n=1,\ldots,N$, to serve as our benchmark truth-level sample. The resulting two dimensional distributions are shown in Fig.~\ref{2d_heat}. 

We observe from Fig.~\ref{2d_heat} that the $N_{j}$ and $N_{b}$ distributions do not appear to be strictly (conditionally) independent. This is evidenced by the fact that different rows (columns) show different bin hierarchies depending on the column (row) they are conditioned on. These effects arise from experimental systematics such as imperfect b-tagging and different (b-)jet acceptances, as well as from statistical fluctuations due to finite sample sizes involved: the sample size of the Monte Carlo simulation and the sample size of the expected events at to the collider luminosity considered. Category bins with very small event yields are particularly affected by these later effects. In Sec~\ref{sec:conditional} we study in detail how well our model approximates the true data distributions even when conditional independence is not exact. We find that for the foreseen (HL)LHC luminosities our model is statistically indistinguishable from the data while retaining classification power to infer the 4-top and $t\overline{t}W^{\pm}$ distributions.

On the other hand, regarding the potential MC mismodelling, we would like to emphasize that our model is aimed to work directly on data and thus address this very kind of problem. That is, we care that our model recovers the true underlying distribution with imperfect (i.e. MC based) priors. In this context we use  MC simulations as stand-in mock data for actual (mixed) distribution measurements  
and apply our model to this mock data with imperfect knowledge encoded in the priors. In order to emulate an imperfect MC prior we skewed the corresponding $N_j$ and $N_b$ distributions from $X$ to higher values and incorporated this into our model through the prior hyperparameters. In general, we can write the hyperparameters $\eta$ of a $V$-dimensional Dirichlet distribution of a random variable $\theta$ as $\eta_{v}=\Sigma\cdot p_{v}$, for $v= 1,\dots,V$. Here $p$ is a multinomial probability distribution and $\Sigma$ is a normalization factor. The role of $p_{v}$ and $\Sigma$ can be understood by looking at the mean and variance of $\theta_{v}$:
%
%
\begin{eqnarray}
    \mathbb{E}[\theta_{v}]&=&p_{v}\nonumber\\
    \text{Var }[\theta_{v}]&=&\frac{p_{v}(1-p_{v})}{\Sigma+1}\,.
\end{eqnarray}

\begin{figure*}[t]
    \centering
    \includegraphics[width=\textwidth]{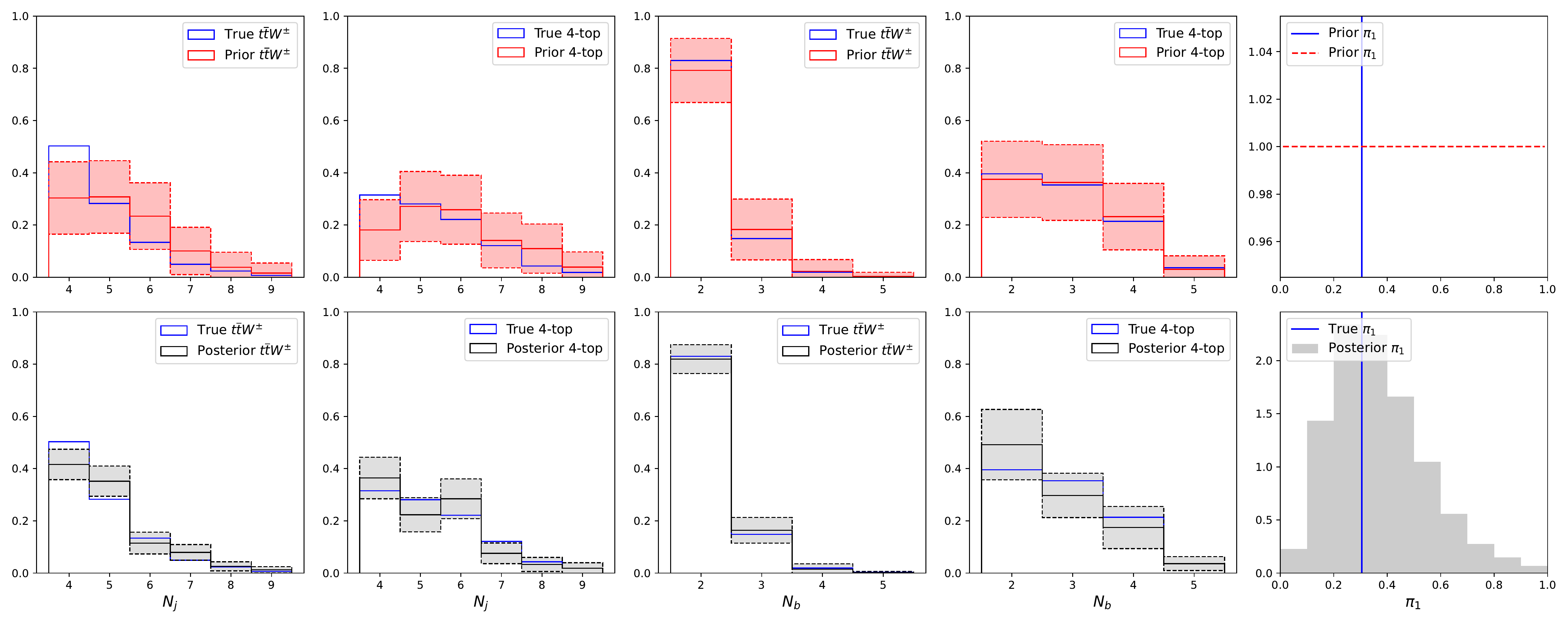}
    \caption{$N_j$, $N_b$ and $\pi_{1}$ distributions: true values (blue), priors (red) and posterior (black).  Shaded regions in first four plots indicate the $1\sigma$ uncertainty region.  Comparing the posteriors to the priors one can appreciate the improvement in estimating the true distributions departing from incorrect and uncertain priors using Bayesian Inference on the data.}
    \label{results}
\end{figure*}

From these equations, we see that $p_{v}$ represents the expected value of $\theta_{v}$ while $\Sigma$ controls the confidence we have on that expectation. We fixed the $p_{v}$ values of the priors for $\alpha$ and $\beta$ in their respective Dirichlets to the normalized $N_{j}$ and $N_{b}$ populations given by the imperfect MC predictions. To reflect our confidence in this estimate, in this example we chose $\Sigma=10$ for each Dirichlet. See Fig.~\ref{results} upper row, where we plot the central values and $1\sigma$ ranges for the prior distributions for $\alpha$ and $\beta$. In an actual experimental analysis, $\Sigma$ could be chosen such that the priors cover all reasonable ranges of the modeled observables. As an extreme example, for the prior on the $\pi$ parameter, giving the fraction of signal and background in the sample, we take a uniform distribution, indicating no prior knowledge on how much background and signal we can expect in the dataset.

\begin{figure}
    \centering
    \includegraphics[width=0.5\textwidth]{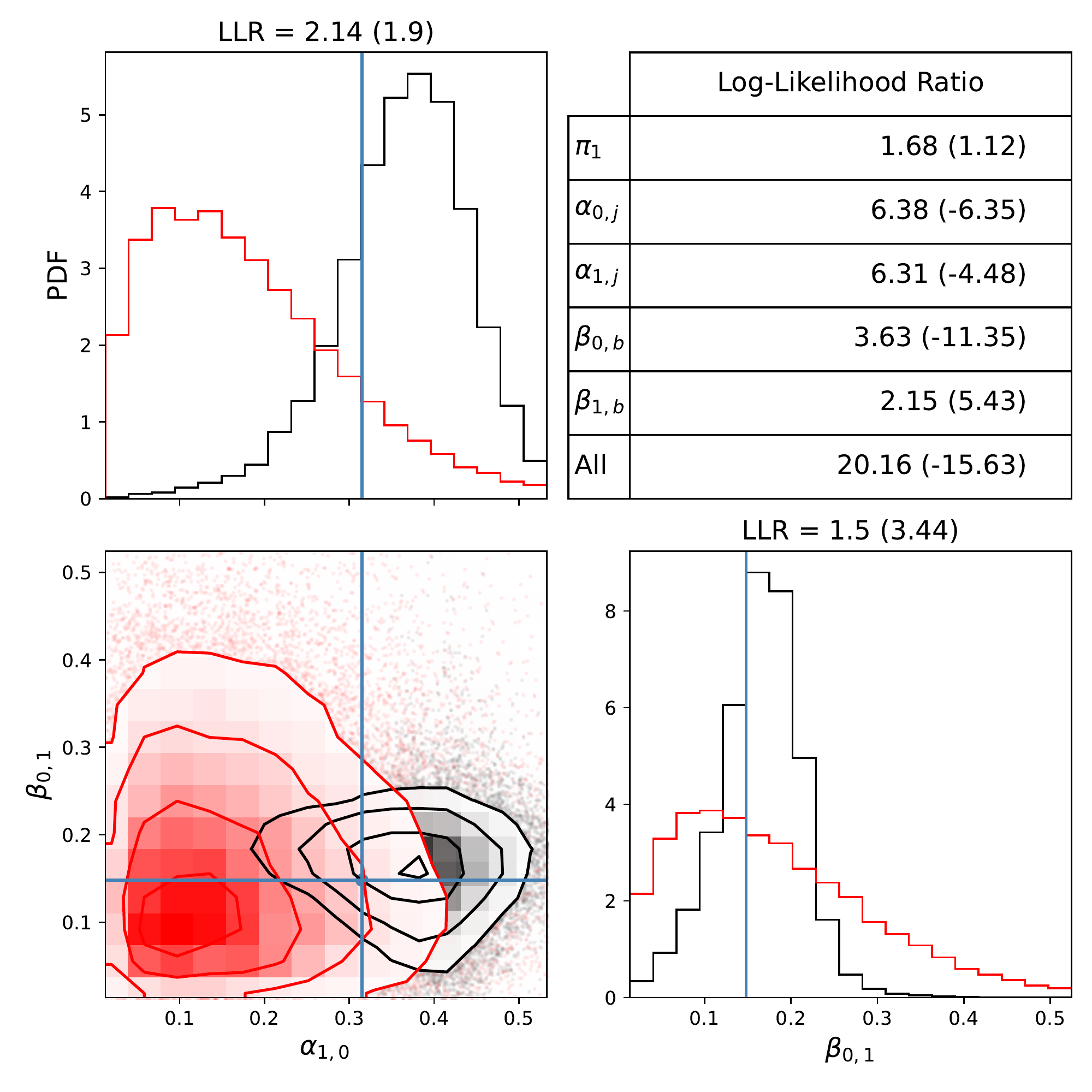}
    \caption{Excerpt from the corner-plot.  Red indicates the prior distribution, black the posterior distribution obtained through GS and blue is the true value. We see how the posterior distribution captures the correlation between $N_{j}$ and $N_{b}$. The titles of each 1D histogram contain the Log-Likelihood Ratio between the posterior and the prior using either GS or VI for the posterior estimation, with the latter shown in parentheses. The table contains the sum of Log-Likelihood Ratios per parameter block, again considering the posteriors obtained through GS and through VI. We see that VI is a bad approximation to GS, failing to improve on the prior for several parameter blocks.}
    \label{corner}
\end{figure}

As we do for the toy model, we study the posterior distribution obtained using GS through the corner-plot, with its LLR partial and global and sums, and through histograms that condense the class-$0$ and class-$1$ $N_{j}$ and $N_{b}$ probability distributions and the $\pi$ probability distribution. We show an excerpt of the corner-plot in Fig.~\ref{corner}. The global sum of the LLRs is $\approx$ 20, reflecting an improvement over the prior. In comparison, the VI estimated posterior does not show an improvement over the prior. This is due to the narrow width of the approximation which excludes the true values of the parameters to a higher level than the more accurate GS obtained posterior estimation. 

In Fig.~\ref{results} we show the results for $N_j$ and $N_b$ distributions of the signal and background, as well as for the $\pi$ parameter, i.e. the fraction of signal in the sample. As in the toy model case, the posterior exhibits good convergence to the true values as well as a considerable reduction of the uncertainty when compared to the prior which emulates the imperfect MC. However, in this case the improvement is different for each feature. The $N_j$ distribution shows a larger improvement, as expected from Fig.~\ref{corner}, while the $N_{b}$ distribution is harder to reconstruct due to the much larger fraction of events populating the first bin. Similar results are obtained for other cases which differ in signal-to-background ratio and number of events. It is also interesting to notice in Fig.~\ref{results} how again from a complete ignorance of the signal and background fractions in the 2LSS++ sample the algorithm recovers a pdf for $\pi$ in good agreement with its true value. We also checked that this agreement holds for other truth values of $\pi$, and that the matching only worsens as the value of $\pi$ approaches the boundaries of $[0,1]$.

In summary, we find that the algorithm successfully infers the $N_j$ and $N_b$ distributions as well as the signal/background fractions. Notably, the best inference occurs for the $N_j$ distribution, which is usually the hardest to predict correctly through MC simulations based on perturbative QCD calculations matched to parton shower algorithms. 

\section{Testing model validity}\label{sec:conditional}

Our method hinges on the validity of the underlying statistical (generative) model. Thus it is imperative to understand how well our model that assumes conditional independence, approximates the true data distributions even when their conditional independence is not exact. To quantify the agreement between the data and our model we consider the mutual information (MI) $I(N_{j},N_{b})$ between $N_{j}$ and $N_{b}$, 
\begin{eqnarray}
    I(N_{j},N_{b}) &=& D_{\text{KL}}(p({j},{b})||p({j})p({b})) \\
    &=& \sum_{{j}=4}^{9} \sum_{{b}=2}^{5}p({j},{b})\text{ Ln }\frac{p({j},{b})}{p({j})p({b})}\,. \nonumber
    \label{eqs_metrics}
\end{eqnarray}
The MI encodes how much information is lost by approximating the full distribution with the product of the two marginal distributions. We can also condition the MI on the class label and obtain the MI for each process $I(N_{j},N_{b}|z)$. By combining the per process MI, we build the conditional MI $I(N_{j},N_{b}|Z)=\sum_{z}p(z)I(N_{j},N_{b}|z)$ which encodes our exact model hypothesis: the data follows a probability distribution which can be written as a combination of two processes, each of which presents a factorized probability distribution. We should note that $I(N_{j},N_{b}|z)$ and $I(N_{j},N_{b}|Z)$ depend explicitly on the availability of labelled data and thus are not computable purely from measured distributions. However, because we expect the simulations to be qualitatively reasonable approximations to measurements, studying the validity of the modelling hypothesis using MC simulations is justified.

Using our finite 4-top and $t\bar t W^\pm$ dataset, we can estimate the relevant probability distributions and obtain finite sample estimations of the relevant MIs. In the large statistics limit, the estimator follows compact asymptotic distributions~\cite{mutual_info_asympt}. However, we are dealing with finite event samples where some category bins are scarcely populated. Thus, in order to quantify the compatibility of our model with the data, we do a series of pseudo-experiments according to the following procedure:
\begin{enumerate}
    \item We take the expected event rates obtained from the {\tt Madgraph}+{\tt Pythia}+{\tt Delphes} pipeline and their uncertainties to generate 2500 pseudo-datasets. For each pseudo-dataset, we sample the expected event rate for each bin according to a Gaussian centered in the MC central value and with the appropriate uncertainty. Then, we sample the observed events for that bin through a Poisson distribution.
    \item For each of these pseudo-datasets, we compute the two-dimensional probability distribution and the marginals for each process and for the full dataset. With these, we obtain the estimators of all four relevant MIs $\hat{I}(N_{j},N_{b}|z)$, with $z= t\overline{t}W^{\pm}, \text{4-top}$, $\hat{I}(N_{j},N_{b})$ and $\hat{I}(N_{j},N_{b}|Z)$.
    \item We use these estimators to study the validity of approximating the joint probability distribution with a certain modelling hypothesis. To this end we construct the probability distribution of the estimator by generating another batch of 2500 pseudo-datasets. This time, each pseudo-dataset is generated using the relevant approximation: for $I(N_{j},N_{b}|z)$, we generate the pseudo-datasets with $p({j}|z)p({b}|z)$; for $I(N_{j},N_{b})$, we generate the pseudo-datasets with $p({j})p({b})$; and for $I(N_{j},N_{b}|Z)$, we generate the pseudo-datasets with $\sum_{z} p(z)p({j}|z)p({b}|z)$. The hypothesis that the obtained estimators $\hat{I}$ are sampled according to the model is the null hypothesis $H_{0}$.
    \item Having obtained the probability distribution of each estimator conditioned on its null hypothesis $H_{0}$ using these additional pseudo-datasets, we compute the one-sided p-value for the "measured" estimator which allows us to discard the null hypothesis with a certain confidence level\footnote{Although not explicit, there is an assumed alternative hypothesis $H_{1}$: the saturated model. For a given pseudo-dataset of $N$ events sampled from a multinomial distribution, its MI is nothing more than $2N$ times its saturated log-likelihood
    ~\cite{BAKER1984437}.}. The p-value can be computed as
    \begin{equation}
        \text{p-value} = \int_{\hat{I}}^{\infty} p(I|H_{0})dI\nonumber
    \end{equation}
    where $I$ can be any of the four metrics considered and $H_{0}$ its associated null hypothesis. In the large statistics limit, this one-sided test asymptotically converges to the compact formulae considered in Ref.~\cite{mutual_info_asympt}.
\end{enumerate}

\begin{figure*}[t]
    \centering
    \includegraphics[width=1.0\textwidth]{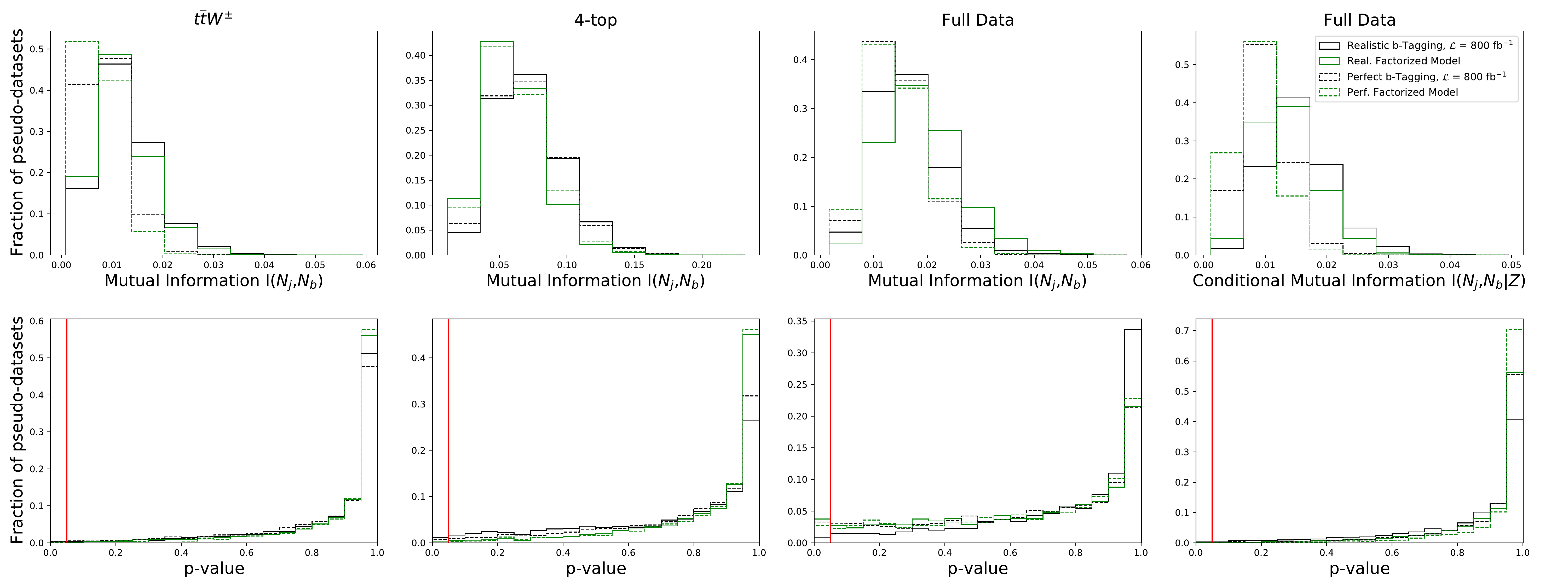}
    \caption{Top row: We show in solid (dashed) black lines the MI between $N_{j}$ and $N_{b}$ with realistic (perfect) b-tagging. In solid (dashed) green we show the MI distribution for the expected event rates which respect conditional independence with realistic (perfect) b-tagging. Bottom row: We show with the same color and line conventions the p-value of the null hypothesis distribution of each estimator. We show in red the $p=0.05$ conventional exclusion value. We can see that for the considered luminosity, $N_{j}$ and $N_{b}$ cannot be ruled out to be conditionally mutually independent. See text for details.}
    \label{mutual_info_analysis}
\end{figure*}

We show the results of this procedure in Fig.~\ref{mutual_info_analysis} for four types of pseudo-datasets. In solid black line we show the pseudo-dataset generated with the expected events as obtained from the {\tt Madgraph}+{\tt Pythia}+{\tt Delphes} pipeline. In dashed black line we consider the event rates we obtain when considering perfect b-tagging in {\tt Delphes}. We do this to verify whether the introduction of imperfect b-tagging, and the resulting correlations between the number of light- and b-jets, spoil conditional independence. In green solid and dashed lines we modify the sampled expected event rates to ensure conditional independence for realistic and perfect b-tagging. These two pseudo-datasets thus agree with our modelling hypothesis and provide a self-consistency check. One should note that the Poisson sampling with relatively small event rates induces a slight violation of conditional independence as it is done in a bin by bin basis. 

In Fig.~\ref{mutual_info_analysis}, we observe that for the considered luminosity $\mathcal L \simeq 800$\,fb${}^{-1}$, the data and our model are not statistically distinguishable from each other. This can be seen from the first, second and fourth columns, where the null hypothesis coincides with the green curves. The p-value distributions in the first and second column imply that 4-top and $t\overline{t}W^{\pm}$ cannot be ruled out to have factorized $(N_{j},N_{b})$ distributions while the fourth column implies the same for the full data and the model which assumes conditional independence. For the third column, both the data and the model are different from the null hypothesis that considers full independence between $N_{j}$ and $N_{b}$. In that case, both the model and the data show slight disagreements with the null hypothesis although they remain compatible with it. We observe how the p-value distribution is more tilted towards the discarded region for the MI compared to the Conditional MI for the full data distribution, specially for perfect b-tagging. Conditional independence is thus a reasonable modelling hypothesis that yields qualitatively different behavior than assuming a single process with a factorized ($N_{j}$,$N_{b}$) distribution. Because conditional independence assumes that correlations between light- and b-jets are induced by marginalizing over the labels, the model acquires classification power for the underlying processes (that we can match to 4-top and $t\overline{t}W^{\pm}$) by learning the induced correlations to achieve explanatory power over the full data distribution.

\begin{figure*}[t]
    \centering
    \includegraphics[width=1.0\textwidth]{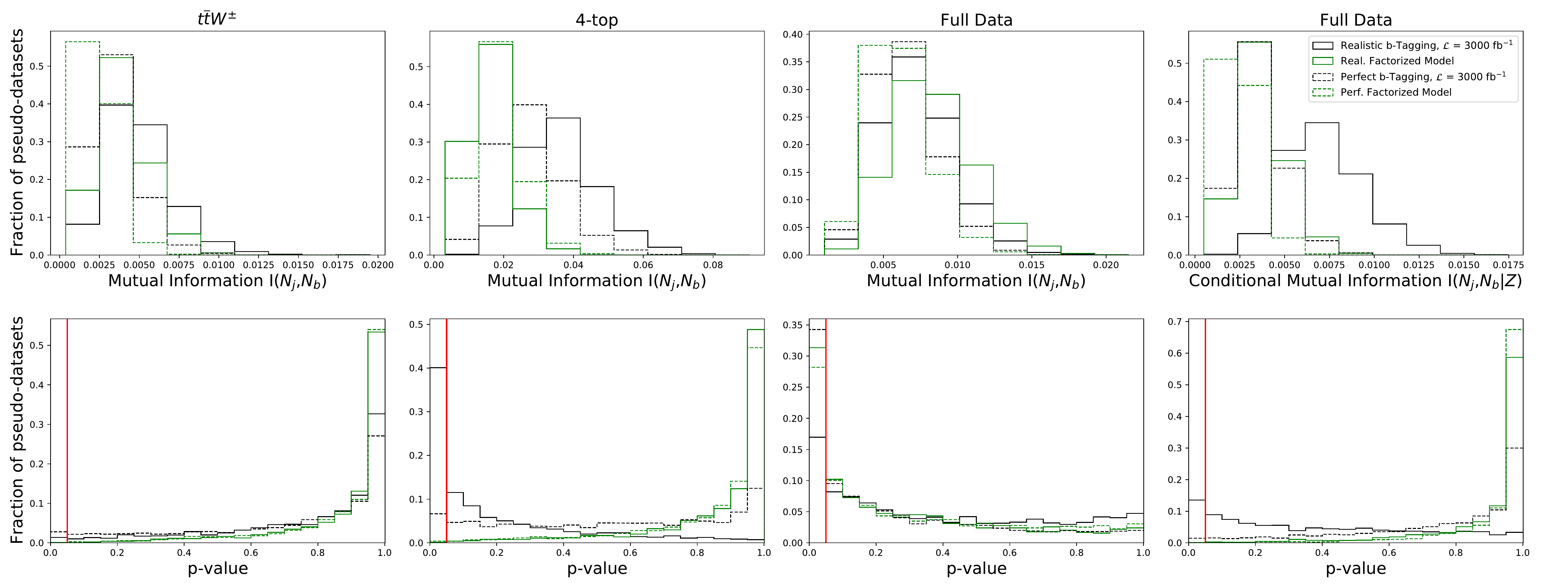}
    \caption{Same as Fig.~\ref{mutual_info_analysis} but for projected High-Luminosity expected event rates.}
    \label{mutual_info_analysis_hl_lhc}
\end{figure*}

The different hypotheses become better distinguishable at larger luminosities. This is seen in Fig.~\ref{mutual_info_analysis_hl_lhc} where we show the results for High-Luminosity LHC projections with $\mathcal{L} = 3000\text{ fb}^{-1}$. 
We observe that 4-top exhibits larger deviations from independence than $t\overline{t}W^{\pm}$. 
In particular $N_{j}$ and $N_{b}$ independence can be ruled out for the 4-top distribution with realistic b-tagging. This in turn causes the full data distribution to be tilted towards lower p-values for the conditionally independent null hypothesis. The $t\overline{t}W^{\pm}$ does not exhibit the same behavior.
We verify that the MI of both processes decreases considerably in the case of (near) perfect b-tagging. 
 In particular, joint (black) 4-top distribution is much closer to its marginalized (green) counterpart which is also reflected in the full data conditional MI distribution
This implies that imperfect b-tagging is indeed an important factor behind observed deviations from the conditional independence hypothesis although it is not the only one. Because we are considering a probabilistic model for the data, a feasible sophistication of this model that includes b-tagging efficiencies as a random variable could restore conditional independence while keeping the number of parameters under control. Such incorporation of the b-tagging efficiency would be the analogue to the introduction of an associated nuisance parameter in traditional statistical analyses.
 Another key feature at HL-LHC luminosity is that for all four pseudo-datasets full independence between $N_{j}$ and $N_{b}$ can be ruled out, as evidenced by the third column. For perfect b-tagging, we can conclude that conditional independence is a valid approximation which yields learnable distributions with discriminatory power between processes. If imperfect b-tagging is taken into account in the generative model then conditional independence remains a valid modelling hypothesis with explanatory power for the full range of luminosities expected at the LHC.

\section{Conclusions}\label{sec:Conclusions}

In summary, we have proposed a new technique to extract signal and background features and fractions relevant for measurements of four-top production at the LHC using Bayesian Inference on the $N_j$ and $N_b$ jet multiplicity distributions. It relies on the assumption of conditional (upon signal and background class) independence of the inferred distributions and harnesses the resulting correlations between $N_j$ and $N_b$ within each class. The algorithm is weakly-supervised since, in addition to data (in the signal region), it only relies on imperfect a priori knowledge how the signal and background differ in their $N_j$ and $N_b$ distributions. Using these results we have proposed a novel approach to test or tune MC predictions in the signal region. Alternatively, it could allow to measure four-top production cross-section and/or test for NP effects in a novel way that alleviates the dependence on MC simulations altogether, as also proposed in Ref.~\cite{Alvarez:2019knh}. 
One could for instance tune the MC in the signal region using the class-$0$ (background) $N_j$ and $N_b$ distributions and then simulate the signal using the tuned MC to check whether its predicted fraction in the 2LSS++ sample agrees with the predictions in $p(\pi|X)$.  Moreover, one can also check whether the MC signal $N_j$ and $N_b$ distributions match the $p(\alpha_{1,i}|X)$ and $p(\beta_{1,i}|X)$ inferred by the algorithm. Using these ideas one would be able effectively to compute acceptances with a MC tuned {\it in-situ} in the signal region, while simultaneously measure the four-top cross-section, or study potential NP contributions to the signal or the backgrounds. 

Certainly, our method as presented in Sec.~\ref{sec:Mixture} is general and applicable also to other particle physics scenarios beside four-top production and potentially opens new venues of searches for NP at colliders. Certainly however, much further work is needed to implement these techniques into feasible experimental analyses. 

\mysection{Acknowledgements} JFK acknowledges the financial support from the Slovenian Research Agency (grant No. J1-3013 and research core funding No. P1-0035). BD acknowledges funding from BMBF. DAF has received funding from the European Research Council (ERC) under the European Union’s Horizon 2020 research and innovation program under grant agreement 833280 (FLAY), and by the Swiss National Science Foundation (SNF) under contract 200021-175940.  We thank the Referee for his/her report, which has considerably improved the content of the article.

\bibliography{current}

\begin{thebibliography}{36}%
\makeatletter
\providecommand \@ifxundefined [1]{%
 \@ifx{#1\undefined}
}%
\providecommand \@ifnum [1]{%
 \ifnum #1\expandafter \@firstoftwo
 \else \expandafter \@secondoftwo
 \fi
}%
\providecommand \@ifx [1]{%
 \ifx #1\expandafter \@firstoftwo
 \else \expandafter \@secondoftwo
 \fi
}%
\providecommand \natexlab [1]{#1}%
\providecommand \enquote  [1]{``#1''}%
\providecommand \bibnamefont  [1]{#1}%
\providecommand \bibfnamefont [1]{#1}%
\providecommand \citenamefont [1]{#1}%
\providecommand \href@noop [0]{\@secondoftwo}%
\providecommand \href [0]{\begingroup \@sanitize@url \@href}%
\providecommand \@href[1]{\@@startlink{#1}\@@href}%
\providecommand \@@href[1]{\endgroup#1\@@endlink}%
\providecommand \@sanitize@url [0]{\catcode `\\12\catcode `\$12\catcode
  `\&12\catcode `\#12\catcode `\^12\catcode `\_12\catcode `\%12\relax}%
\providecommand \@@startlink[1]{}%
\providecommand \@@endlink[0]{}%
\providecommand \url  [0]{\begingroup\@sanitize@url \@url }%
\providecommand \@url [1]{\endgroup\@href {#1}{\urlprefix }}%
\providecommand \urlprefix  [0]{URL }%
\providecommand \Eprint [0]{\href }%
\providecommand \doibase [0]{https://doi.org/}%
\providecommand \selectlanguage [0]{\@gobble}%
\providecommand \bibinfo  [0]{\@secondoftwo}%
\providecommand \bibfield  [0]{\@secondoftwo}%
\providecommand \translation [1]{[#1]}%
\providecommand \BibitemOpen [0]{}%
\providecommand \bibitemStop [0]{}%
\providecommand \bibitemNoStop [0]{.\EOS\space}%
\providecommand \EOS [0]{\spacefactor3000\relax}%
\providecommand \BibitemShut  [1]{\csname bibitem#1\endcsname}%
\let\auto@bib@innerbib\@empty
\bibitem [{\citenamefont {Kasieczka}\ \emph {et~al.}(2021)\citenamefont
  {Kasieczka} \emph {et~al.}}]{Kasieczka:2021xcg}%
  \BibitemOpen
  \bibfield  {author} {\bibinfo {author} {\bibfnamefont {G.}~\bibnamefont
  {Kasieczka}} \emph {et~al.},\ }\bibfield  {title} {\bibinfo {title} {{The LHC
  Olympics 2020: A Community Challenge for Anomaly Detection in High Energy
  Physics}},\ }\href@noop {} {\  (\bibinfo {year} {2021})},\ \Eprint
  {https://arxiv.org/abs/2101.08320} {arXiv:2101.08320 [hep-ph]} \BibitemShut
  {NoStop}%
\bibitem [{\citenamefont {Kasieczka}\ \emph {et~al.}(2020)\citenamefont
  {Kasieczka}, \citenamefont {Nachman}, \citenamefont {Schwartz},\ and\
  \citenamefont {Shih}}]{Kasieczka:2020pil}%
  \BibitemOpen
  \bibfield  {author} {\bibinfo {author} {\bibfnamefont {G.}~\bibnamefont
  {Kasieczka}}, \bibinfo {author} {\bibfnamefont {B.}~\bibnamefont {Nachman}},
  \bibinfo {author} {\bibfnamefont {M.~D.}\ \bibnamefont {Schwartz}},\ and\
  \bibinfo {author} {\bibfnamefont {D.}~\bibnamefont {Shih}},\ }\bibfield
  {title} {\bibinfo {title} {{ABCDisCo: Automating the ABCD Method with Machine
  Learning}},\ }\href@noop {} {\  (\bibinfo {year} {2020})},\ \Eprint
  {https://arxiv.org/abs/2007.14400} {arXiv:2007.14400 [hep-ph]} \BibitemShut
  {NoStop}%
\bibitem [{\citenamefont {Ghosh}\ \emph {et~al.}(2021)\citenamefont {Ghosh},
  \citenamefont {Nachman},\ and\ \citenamefont {Whiteson}}]{Ghosh:2021roe}%
  \BibitemOpen
  \bibfield  {author} {\bibinfo {author} {\bibfnamefont {A.}~\bibnamefont
  {Ghosh}}, \bibinfo {author} {\bibfnamefont {B.}~\bibnamefont {Nachman}},\
  and\ \bibinfo {author} {\bibfnamefont {D.}~\bibnamefont {Whiteson}},\
  }\bibfield  {title} {\bibinfo {title} {{Uncertainty Aware Learning for High
  Energy Physics}},\ }\href@noop {} {\  (\bibinfo {year} {2021})},\ \Eprint
  {https://arxiv.org/abs/2105.08742} {arXiv:2105.08742 [physics.data-an]}
  \BibitemShut {NoStop}%
\bibitem [{\citenamefont {Benkendorfer}\ \emph {et~al.}(2020)\citenamefont
  {Benkendorfer}, \citenamefont {Pottier},\ and\ \citenamefont
  {Nachman}}]{Benkendorfer:2020gek}%
  \BibitemOpen
  \bibfield  {author} {\bibinfo {author} {\bibfnamefont {K.}~\bibnamefont
  {Benkendorfer}}, \bibinfo {author} {\bibfnamefont {L.~L.}\ \bibnamefont
  {Pottier}},\ and\ \bibinfo {author} {\bibfnamefont {B.}~\bibnamefont
  {Nachman}},\ }\bibfield  {title} {\bibinfo {title} {{Simulation-Assisted
  Decorrelation for Resonant Anomaly Detection}},\ }\href@noop {} {\  (\bibinfo
  {year} {2020})},\ \Eprint {https://arxiv.org/abs/2009.02205}
  {arXiv:2009.02205 [hep-ph]} \BibitemShut {NoStop}%
\bibitem [{\citenamefont {Choi}\ \emph {et~al.}(2020)\citenamefont {Choi},
  \citenamefont {Lim},\ and\ \citenamefont {Oh}}]{Choi:2020bnf}%
  \BibitemOpen
  \bibfield  {author} {\bibinfo {author} {\bibfnamefont {S.}~\bibnamefont
  {Choi}}, \bibinfo {author} {\bibfnamefont {J.}~\bibnamefont {Lim}},\ and\
  \bibinfo {author} {\bibfnamefont {H.}~\bibnamefont {Oh}},\ }\bibfield
  {title} {\bibinfo {title} {{Data-driven Estimation of Background Distribution
  through Neural Autoregressive Flows}},\ }\href@noop {} {\  (\bibinfo {year}
  {2020})},\ \Eprint {https://arxiv.org/abs/2008.03636} {arXiv:2008.03636
  [hep-ph]} \BibitemShut {NoStop}%
\bibitem [{\citenamefont {Flesher}\ \emph {et~al.}(2020)\citenamefont
  {Flesher}, \citenamefont {Fraser}, \citenamefont {Hutchison}, \citenamefont
  {Ostdiek},\ and\ \citenamefont {Schwartz}}]{Flesher:2020kuy}%
  \BibitemOpen
  \bibfield  {author} {\bibinfo {author} {\bibfnamefont {F.}~\bibnamefont
  {Flesher}}, \bibinfo {author} {\bibfnamefont {K.}~\bibnamefont {Fraser}},
  \bibinfo {author} {\bibfnamefont {C.}~\bibnamefont {Hutchison}}, \bibinfo
  {author} {\bibfnamefont {B.}~\bibnamefont {Ostdiek}},\ and\ \bibinfo {author}
  {\bibfnamefont {M.~D.}\ \bibnamefont {Schwartz}},\ }\bibfield  {title}
  {\bibinfo {title} {{Parameter Inference from Event Ensembles and the
  Top-Quark Mass}},\ }\href@noop {} {\  (\bibinfo {year} {2020})},\ \Eprint
  {https://arxiv.org/abs/2011.04666} {arXiv:2011.04666 [hep-ph]} \BibitemShut
  {NoStop}%
\bibitem [{\citenamefont {Lillie}\ \emph {et~al.}(2008)\citenamefont {Lillie},
  \citenamefont {Shu},\ and\ \citenamefont {Tait}}]{Lillie:2007hd}%
  \BibitemOpen
  \bibfield  {author} {\bibinfo {author} {\bibfnamefont {B.}~\bibnamefont
  {Lillie}}, \bibinfo {author} {\bibfnamefont {J.}~\bibnamefont {Shu}},\ and\
  \bibinfo {author} {\bibfnamefont {T.~M.~P.}\ \bibnamefont {Tait}},\
  }\bibfield  {title} {\bibinfo {title} {{Top Compositeness at the Tevatron and
  LHC}},\ }\href {https://doi.org/10.1088/1126-6708/2008/04/087} {\bibfield
  {journal} {\bibinfo  {journal} {JHEP}\ }\textbf {\bibinfo {volume} {04}},\
  \bibinfo {pages} {087}},\ \Eprint {https://arxiv.org/abs/0712.3057}
  {arXiv:0712.3057 [hep-ph]} \BibitemShut {NoStop}%
\bibitem [{\citenamefont {Kumar}\ \emph {et~al.}(2009)\citenamefont {Kumar},
  \citenamefont {Tait},\ and\ \citenamefont {Vega-Morales}}]{Kumar:2009vs}%
  \BibitemOpen
  \bibfield  {author} {\bibinfo {author} {\bibfnamefont {K.}~\bibnamefont
  {Kumar}}, \bibinfo {author} {\bibfnamefont {T.~M.~P.}\ \bibnamefont {Tait}},\
  and\ \bibinfo {author} {\bibfnamefont {R.}~\bibnamefont {Vega-Morales}},\
  }\bibfield  {title} {\bibinfo {title} {{Manifestations of Top Compositeness
  at Colliders}},\ }\href {https://doi.org/10.1088/1126-6708/2009/05/022}
  {\bibfield  {journal} {\bibinfo  {journal} {JHEP}\ }\textbf {\bibinfo
  {volume} {05}},\ \bibinfo {pages} {022}},\ \Eprint
  {https://arxiv.org/abs/0901.3808} {arXiv:0901.3808 [hep-ph]} \BibitemShut
  {NoStop}%
\bibitem [{\citenamefont {Acharya}\ \emph {et~al.}(2009)\citenamefont
  {Acharya}, \citenamefont {Grajek}, \citenamefont {Kane}, \citenamefont
  {Kuflik}, \citenamefont {Suruliz},\ and\ \citenamefont
  {Wang}}]{Acharya:2009gb}%
  \BibitemOpen
  \bibfield  {author} {\bibinfo {author} {\bibfnamefont {B.~S.}\ \bibnamefont
  {Acharya}}, \bibinfo {author} {\bibfnamefont {P.}~\bibnamefont {Grajek}},
  \bibinfo {author} {\bibfnamefont {G.~L.}\ \bibnamefont {Kane}}, \bibinfo
  {author} {\bibfnamefont {E.}~\bibnamefont {Kuflik}}, \bibinfo {author}
  {\bibfnamefont {K.}~\bibnamefont {Suruliz}},\ and\ \bibinfo {author}
  {\bibfnamefont {L.-T.}\ \bibnamefont {Wang}},\ }\bibfield  {title} {\bibinfo
  {title} {{Identifying Multi-Top Events from Gluino Decay at the LHC}},\
  }\href@noop {} {\  (\bibinfo {year} {2009})},\ \Eprint
  {https://arxiv.org/abs/0901.3367} {arXiv:0901.3367 [hep-ph]} \BibitemShut
  {NoStop}%
\bibitem [{\citenamefont {Kim}\ \emph {et~al.}(2016)\citenamefont {Kim},
  \citenamefont {Kong}, \citenamefont {Lee},\ and\ \citenamefont
  {Mohlabeng}}]{Kim:2016plm}%
  \BibitemOpen
  \bibfield  {author} {\bibinfo {author} {\bibfnamefont {J.~H.}\ \bibnamefont
  {Kim}}, \bibinfo {author} {\bibfnamefont {K.}~\bibnamefont {Kong}}, \bibinfo
  {author} {\bibfnamefont {S.~J.}\ \bibnamefont {Lee}},\ and\ \bibinfo {author}
  {\bibfnamefont {G.}~\bibnamefont {Mohlabeng}},\ }\bibfield  {title} {\bibinfo
  {title} {{Probing TeV scale Top-Philic Resonances with Boosted Top-Tagging at
  the High Luminosity LHC}},\ }\href
  {https://doi.org/10.1103/PhysRevD.94.035023} {\bibfield  {journal} {\bibinfo
  {journal} {Phys. Rev.}\ }\textbf {\bibinfo {volume} {D94}},\ \bibinfo {pages}
  {035023} (\bibinfo {year} {2016})},\ \Eprint
  {https://arxiv.org/abs/1604.07421} {arXiv:1604.07421 [hep-ph]} \BibitemShut
  {NoStop}%
\bibitem [{\citenamefont {Liu}\ and\ \citenamefont
  {Mahbubani}(2016)}]{Liu:2015hxi}%
  \BibitemOpen
  \bibfield  {author} {\bibinfo {author} {\bibfnamefont {D.}~\bibnamefont
  {Liu}}\ and\ \bibinfo {author} {\bibfnamefont {R.}~\bibnamefont
  {Mahbubani}},\ }\bibfield  {title} {\bibinfo {title} {{Probing top-antitop
  resonances with $t\bar{t}$ scattering at LHC14}},\ }\href
  {https://doi.org/10.1007/JHEP04(2016)116} {\bibfield  {journal} {\bibinfo
  {journal} {JHEP}\ }\textbf {\bibinfo {volume} {04}},\ \bibinfo {pages}
  {116}},\ \Eprint {https://arxiv.org/abs/1511.09452} {arXiv:1511.09452
  [hep-ph]} \BibitemShut {NoStop}%
\bibitem [{\citenamefont {Aguilar-Saavedra}\ and\ \citenamefont
  {Santiago}(2012)}]{AguilarSaavedra:2011ck}%
  \BibitemOpen
  \bibfield  {author} {\bibinfo {author} {\bibfnamefont {J.~A.}\ \bibnamefont
  {Aguilar-Saavedra}}\ and\ \bibinfo {author} {\bibfnamefont {J.}~\bibnamefont
  {Santiago}},\ }\bibfield  {title} {\bibinfo {title} {{Four tops and the $t
  \bar{t}$ forward-backward asymmetry}},\ }\href
  {https://doi.org/10.1103/PhysRevD.85.034021} {\bibfield  {journal} {\bibinfo
  {journal} {Phys. Rev.}\ }\textbf {\bibinfo {volume} {D85}},\ \bibinfo {pages}
  {034021} (\bibinfo {year} {2012})},\ \Eprint
  {https://arxiv.org/abs/1112.3778} {arXiv:1112.3778 [hep-ph]} \BibitemShut
  {NoStop}%
\bibitem [{\citenamefont {Camargo-Molina}\ \emph {et~al.}(2018)\citenamefont
  {Camargo-Molina}, \citenamefont {Celis},\ and\ \citenamefont
  {Faroughy}}]{Camargo-Molina:2018cwu}%
  \BibitemOpen
  \bibfield  {author} {\bibinfo {author} {\bibfnamefont {J.~E.}\ \bibnamefont
  {Camargo-Molina}}, \bibinfo {author} {\bibfnamefont {A.}~\bibnamefont
  {Celis}},\ and\ \bibinfo {author} {\bibfnamefont {D.~A.}\ \bibnamefont
  {Faroughy}},\ }\bibfield  {title} {\bibinfo {title} {{Anomalies in Bottom
  from new physics in Top}},\ }\href
  {https://doi.org/10.1016/j.physletb.2018.07.051} {\bibfield  {journal}
  {\bibinfo  {journal} {Phys. Lett. B}\ }\textbf {\bibinfo {volume} {784}},\
  \bibinfo {pages} {284} (\bibinfo {year} {2018})},\ \Eprint
  {https://arxiv.org/abs/1805.04917} {arXiv:1805.04917 [hep-ph]} \BibitemShut
  {NoStop}%
\bibitem [{\citenamefont {Alvarez}\ \emph {et~al.}(2019)\citenamefont
  {Alvarez}, \citenamefont {Juste},\ and\ \citenamefont
  {Seoane}}]{Alvarez:2019uxp}%
  \BibitemOpen
  \bibfield  {author} {\bibinfo {author} {\bibfnamefont {E.}~\bibnamefont
  {Alvarez}}, \bibinfo {author} {\bibfnamefont {A.}~\bibnamefont {Juste}},\
  and\ \bibinfo {author} {\bibfnamefont {R.~M.~S.}\ \bibnamefont {Seoane}},\
  }\bibfield  {title} {\bibinfo {title} {{Four-top as probe of light top-philic
  New Physics}},\ }\href {https://doi.org/10.1007/JHEP12(2019)080} {\bibfield
  {journal} {\bibinfo  {journal} {JHEP}\ }\textbf {\bibinfo {volume} {12}},\
  \bibinfo {pages} {080}},\ \Eprint {https://arxiv.org/abs/1910.09581}
  {arXiv:1910.09581 [hep-ph]} \BibitemShut {NoStop}%
\bibitem [{\citenamefont {Darm\'e}\ \emph {et~al.}(2021)\citenamefont
  {Darm\'e}, \citenamefont {Fuks},\ and\ \citenamefont
  {Maltoni}}]{Darme:2021xxu}%
  \BibitemOpen
  \bibfield  {author} {\bibinfo {author} {\bibfnamefont {L.}~\bibnamefont
  {Darm\'e}}, \bibinfo {author} {\bibfnamefont {B.}~\bibnamefont {Fuks}},\ and\
  \bibinfo {author} {\bibfnamefont {F.}~\bibnamefont {Maltoni}},\ }\bibfield
  {title} {\bibinfo {title} {{Top-philic heavy resonances in four-top final
  states and their EFT interpretation}},\ }\href@noop {} {\  (\bibinfo {year}
  {2021})},\ \Eprint {https://arxiv.org/abs/2104.09512} {arXiv:2104.09512
  [hep-ph]} \BibitemShut {NoStop}%
\bibitem [{\citenamefont {Khatibi}\ and\ \citenamefont
  {Khanpour}(2021)}]{Khatibi:2020mvt}%
  \BibitemOpen
  \bibfield  {author} {\bibinfo {author} {\bibfnamefont {S.}~\bibnamefont
  {Khatibi}}\ and\ \bibinfo {author} {\bibfnamefont {H.}~\bibnamefont
  {Khanpour}},\ }\bibfield  {title} {\bibinfo {title} {{Probing four-fermion
  operators in the triple top production at future hadron colliders}},\ }\href
  {https://doi.org/10.1016/j.nuclphysb.2021.115432} {\bibfield  {journal}
  {\bibinfo  {journal} {Nucl. Phys. B}\ }\textbf {\bibinfo {volume} {967}},\
  \bibinfo {pages} {115432} (\bibinfo {year} {2021})},\ \Eprint
  {https://arxiv.org/abs/2011.15060} {arXiv:2011.15060 [hep-ph]} \BibitemShut
  {NoStop}%
\bibitem [{\citenamefont {Banelli}\ \emph {et~al.}(2021)\citenamefont
  {Banelli}, \citenamefont {Salvioni}, \citenamefont {Serra}, \citenamefont
  {Theil},\ and\ \citenamefont {Weiler}}]{Banelli:2020iau}%
  \BibitemOpen
  \bibfield  {author} {\bibinfo {author} {\bibfnamefont {G.}~\bibnamefont
  {Banelli}}, \bibinfo {author} {\bibfnamefont {E.}~\bibnamefont {Salvioni}},
  \bibinfo {author} {\bibfnamefont {J.}~\bibnamefont {Serra}}, \bibinfo
  {author} {\bibfnamefont {T.}~\bibnamefont {Theil}},\ and\ \bibinfo {author}
  {\bibfnamefont {A.}~\bibnamefont {Weiler}},\ }\bibfield  {title} {\bibinfo
  {title} {{The Present and Future of Four Top Operators}},\ }\href
  {https://doi.org/10.1007/JHEP02(2021)043} {\bibfield  {journal} {\bibinfo
  {journal} {JHEP}\ }\textbf {\bibinfo {volume} {02}},\ \bibinfo {pages}
  {043}},\ \Eprint {https://arxiv.org/abs/2010.05915} {arXiv:2010.05915
  [hep-ph]} \BibitemShut {NoStop}%
\bibitem [{\citenamefont {Cao}\ \emph {et~al.}(2021)\citenamefont {Cao},
  \citenamefont {Fu}, \citenamefont {Liu}, \citenamefont {Wang},\ and\
  \citenamefont {Zhang}}]{Cao:2021qqt}%
  \BibitemOpen
  \bibfield  {author} {\bibinfo {author} {\bibfnamefont {Q.-H.}\ \bibnamefont
  {Cao}}, \bibinfo {author} {\bibfnamefont {J.-N.}\ \bibnamefont {Fu}},
  \bibinfo {author} {\bibfnamefont {Y.}~\bibnamefont {Liu}}, \bibinfo {author}
  {\bibfnamefont {X.-H.}\ \bibnamefont {Wang}},\ and\ \bibinfo {author}
  {\bibfnamefont {R.}~\bibnamefont {Zhang}},\ }\bibfield  {title} {\bibinfo
  {title} {{Probing Top-philic New Physics via Four-Top-Quark Production}},\
  }\href@noop {} {\  (\bibinfo {year} {2021})},\ \Eprint
  {https://arxiv.org/abs/2105.03372} {arXiv:2105.03372 [hep-ph]} \BibitemShut
  {NoStop}%
\bibitem [{\citenamefont {Alvarez}\ \emph {et~al.}(2020)\citenamefont
  {Alvarez}, \citenamefont {Lamagna},\ and\ \citenamefont
  {Szewc}}]{Alvarez:2019knh}%
  \BibitemOpen
  \bibfield  {author} {\bibinfo {author} {\bibfnamefont {E.}~\bibnamefont
  {Alvarez}}, \bibinfo {author} {\bibfnamefont {F.}~\bibnamefont {Lamagna}},\
  and\ \bibinfo {author} {\bibfnamefont {M.}~\bibnamefont {Szewc}},\ }\bibfield
   {title} {\bibinfo {title} {{Topic Model for four-top at the LHC}},\ }\href
  {https://doi.org/10.1007/JHEP01(2020)049} {\bibfield  {journal} {\bibinfo
  {journal} {JHEP}\ }\textbf {\bibinfo {volume} {01}},\ \bibinfo {pages}
  {049}},\ \bibinfo {note} {[JHEP20,049(2020)]},\ \Eprint
  {https://arxiv.org/abs/1911.09699} {arXiv:1911.09699 [hep-ph]} \BibitemShut
  {NoStop}%
\bibitem [{\citenamefont {Alvarez}\ \emph {et~al.}(2017)\citenamefont
  {Alvarez}, \citenamefont {Faroughy}, \citenamefont {Kamenik}, \citenamefont
  {Morales},\ and\ \citenamefont {Szynkman}}]{Alvarez:2016nrz}%
  \BibitemOpen
  \bibfield  {author} {\bibinfo {author} {\bibfnamefont {E.}~\bibnamefont
  {Alvarez}}, \bibinfo {author} {\bibfnamefont {D.~A.}\ \bibnamefont
  {Faroughy}}, \bibinfo {author} {\bibfnamefont {J.~F.}\ \bibnamefont
  {Kamenik}}, \bibinfo {author} {\bibfnamefont {R.}~\bibnamefont {Morales}},\
  and\ \bibinfo {author} {\bibfnamefont {A.}~\bibnamefont {Szynkman}},\
  }\bibfield  {title} {\bibinfo {title} {{Four Tops for LHC}},\ }\href
  {https://doi.org/10.1016/j.nuclphysb.2016.11.024} {\bibfield  {journal}
  {\bibinfo  {journal} {Nucl. Phys. B}\ }\textbf {\bibinfo {volume} {915}},\
  \bibinfo {pages} {19} (\bibinfo {year} {2017})},\ \Eprint
  {https://arxiv.org/abs/1611.05032} {arXiv:1611.05032 [hep-ph]} \BibitemShut
  {NoStop}%
\bibitem [{\citenamefont {Aad}\ \emph {et~al.}(2021)\citenamefont {Aad} \emph
  {et~al.}}]{ATLAS:2021kqb}%
  \BibitemOpen
  \bibfield  {author} {\bibinfo {author} {\bibfnamefont {G.}~\bibnamefont
  {Aad}} \emph {et~al.} (\bibinfo {collaboration} {ATLAS}),\ }\bibfield
  {title} {\bibinfo {title} {{Measurement of the $t\bar{t}t\bar{t}$ production
  cross section in $pp$ collisions at $\sqrt{s}$=13 TeV with the ATLAS
  detector}},\ }\href@noop {} {\  (\bibinfo {year} {2021})},\ \Eprint
  {https://arxiv.org/abs/2106.11683} {arXiv:2106.11683 [hep-ex]} \BibitemShut
  {NoStop}%
\bibitem [{\citenamefont {Sirunyan}\ \emph {et~al.}(2019)\citenamefont
  {Sirunyan} \emph {et~al.}}]{CMS:2019jsc}%
  \BibitemOpen
  \bibfield  {author} {\bibinfo {author} {\bibfnamefont {A.~M.}\ \bibnamefont
  {Sirunyan}} \emph {et~al.} (\bibinfo {collaboration} {CMS}),\ }\bibfield
  {title} {\bibinfo {title} {{Search for the production of four top quarks in
  the single-lepton and opposite-sign dilepton final states in proton-proton
  collisions at $ \sqrt{s} $ = 13 TeV}},\ }\href
  {https://doi.org/10.1007/JHEP11(2019)082} {\bibfield  {journal} {\bibinfo
  {journal} {JHEP}\ }\textbf {\bibinfo {volume} {11}},\ \bibinfo {pages}
  {082}},\ \Eprint {https://arxiv.org/abs/1906.02805} {arXiv:1906.02805
  [hep-ex]} \BibitemShut {NoStop}%
\bibitem [{\citenamefont {Sirunyan}\ \emph {et~al.}(2020)\citenamefont
  {Sirunyan} \emph {et~al.}}]{Sirunyan:2019wxt}%
  \BibitemOpen
  \bibfield  {author} {\bibinfo {author} {\bibfnamefont {A.~M.}\ \bibnamefont
  {Sirunyan}} \emph {et~al.} (\bibinfo {collaboration} {CMS}),\ }\bibfield
  {title} {\bibinfo {title} {{Search for production of four top quarks in final
  states with same-sign or multiple leptons in proton-proton collisions at
  $\sqrt{s}=$ 13 TeV}},\ }\href
  {https://doi.org/10.1140/epjc/s10052-019-7593-7} {\bibfield  {journal}
  {\bibinfo  {journal} {Eur. Phys. J. C}\ }\textbf {\bibinfo {volume} {80}},\
  \bibinfo {pages} {75} (\bibinfo {year} {2020})},\ \Eprint
  {https://arxiv.org/abs/1908.06463} {arXiv:1908.06463 [hep-ex]} \BibitemShut
  {NoStop}%
\bibitem [{\citenamefont {Aad}\ \emph {et~al.}(2020)\citenamefont {Aad} \emph
  {et~al.}}]{Aad:2020klt}%
  \BibitemOpen
  \bibfield  {author} {\bibinfo {author} {\bibfnamefont {G.}~\bibnamefont
  {Aad}} \emph {et~al.} (\bibinfo {collaboration} {ATLAS}),\ }\bibfield
  {title} {\bibinfo {title} {{Evidence for $t\bar{t}t\bar{t}$ production in the
  multilepton final state in proton\textendash{}proton collisions at
  $\sqrt{s}=13$ $\text {TeV}$ with the ATLAS detector}},\ }\href
  {https://doi.org/10.1140/epjc/s10052-020-08509-3} {\bibfield  {journal}
  {\bibinfo  {journal} {Eur. Phys. J. C}\ }\textbf {\bibinfo {volume} {80}},\
  \bibinfo {pages} {1085} (\bibinfo {year} {2020})},\ \Eprint
  {https://arxiv.org/abs/2007.14858} {arXiv:2007.14858 [hep-ex]} \BibitemShut
  {NoStop}%
\bibitem [{\citenamefont {Dillon}\ \emph {et~al.}(2019)\citenamefont {Dillon},
  \citenamefont {Faroughy},\ and\ \citenamefont {Kamenik}}]{Dillon:2019cqt}%
  \BibitemOpen
  \bibfield  {author} {\bibinfo {author} {\bibfnamefont {B.~M.}\ \bibnamefont
  {Dillon}}, \bibinfo {author} {\bibfnamefont {D.~A.}\ \bibnamefont
  {Faroughy}},\ and\ \bibinfo {author} {\bibfnamefont {J.~F.}\ \bibnamefont
  {Kamenik}},\ }\bibfield  {title} {\bibinfo {title} {{Uncovering latent jet
  substructure}},\ }\href {https://doi.org/10.1103/PhysRevD.100.056002}
  {\bibfield  {journal} {\bibinfo  {journal} {Phys. Rev.}\ }\textbf {\bibinfo
  {volume} {D100}},\ \bibinfo {pages} {056002} (\bibinfo {year} {2019})},\
  \Eprint {https://arxiv.org/abs/1904.04200} {arXiv:1904.04200 [hep-ph]}
  \BibitemShut {NoStop}%
\bibitem [{\citenamefont {Dillon}\ \emph {et~al.}(2020)\citenamefont {Dillon},
  \citenamefont {Faroughy}, \citenamefont {Kamenik},\ and\ \citenamefont
  {Szewc}}]{Dillon:2020quc}%
  \BibitemOpen
  \bibfield  {author} {\bibinfo {author} {\bibfnamefont {B.~M.}\ \bibnamefont
  {Dillon}}, \bibinfo {author} {\bibfnamefont {D.~A.}\ \bibnamefont
  {Faroughy}}, \bibinfo {author} {\bibfnamefont {J.~F.}\ \bibnamefont
  {Kamenik}},\ and\ \bibinfo {author} {\bibfnamefont {M.}~\bibnamefont
  {Szewc}},\ }\bibfield  {title} {\bibinfo {title} {{Learning the latent
  structure of collider events}},\ }\href
  {https://doi.org/10.1007/JHEP10(2020)206} {\bibfield  {journal} {\bibinfo
  {journal} {JHEP}\ }\textbf {\bibinfo {volume} {10}},\ \bibinfo {pages}
  {206}},\ \Eprint {https://arxiv.org/abs/2005.12319} {arXiv:2005.12319
  [hep-ph]} \BibitemShut {NoStop}%
\bibitem [{\citenamefont {Dillon}\ \emph {et~al.}(2021)\citenamefont {Dillon},
  \citenamefont {Plehn}, \citenamefont {Sauer},\ and\ \citenamefont
  {Sorrenson}}]{Dillon:2021nxw}%
  \BibitemOpen
  \bibfield  {author} {\bibinfo {author} {\bibfnamefont {B.~M.}\ \bibnamefont
  {Dillon}}, \bibinfo {author} {\bibfnamefont {T.}~\bibnamefont {Plehn}},
  \bibinfo {author} {\bibfnamefont {C.}~\bibnamefont {Sauer}},\ and\ \bibinfo
  {author} {\bibfnamefont {P.}~\bibnamefont {Sorrenson}},\ }\bibfield  {title}
  {\bibinfo {title} {{Better Latent Spaces for Better Autoencoders}},\
  }\href@noop {} {\  (\bibinfo {year} {2021})},\ \Eprint
  {https://arxiv.org/abs/2104.08291} {arXiv:2104.08291 [hep-ph]} \BibitemShut
  {NoStop}%
\bibitem [{\citenamefont {Bishop}(2006)}]{Bishop:998831}%
  \BibitemOpen
  \bibfield  {author} {\bibinfo {author} {\bibfnamefont {C.~M.}\ \bibnamefont
  {Bishop}},\ }\href {https://cds.cern.ch/record/998831} {\emph {\bibinfo
  {title} {{Pattern recognition and machine learning}}}},\ Information science
  and statistics\ (\bibinfo  {publisher} {Springer},\ \bibinfo {address} {New
  York, NY},\ \bibinfo {year} {2006})\ \bibinfo {note} {softcover published in
  2016}\BibitemShut {NoStop}%
\bibitem [{cod(2021)}]{code-bayes-four-tops}%
  \BibitemOpen
  \href@noop {} {\bibinfo {title} {Bayesian inference for four tops at the
  lhc}},\ \bibinfo {howpublished}
  {\url{https://github.com/ManuelSzewc/bayes-4tops}} (\bibinfo {year}
  {2021})\BibitemShut {NoStop}%
\bibitem [{\citenamefont {Sokal}(1996)}]{Sokal1996MonteCM}%
  \BibitemOpen
  \bibfield  {author} {\bibinfo {author} {\bibfnamefont {A.}~\bibnamefont
  {Sokal}},\ }\bibfield  {title} {\bibinfo {title} {Monte carlo methods in
  statistical mechanics: Foundations and new algorithms note to the reader}\
  }(\bibinfo {year} {1996})\BibitemShut {NoStop}%
\bibitem [{\citenamefont {Foreman-Mackey}\ \emph {et~al.}(2013)\citenamefont
  {Foreman-Mackey}, \citenamefont {Hogg}, \citenamefont {Lang},\ and\
  \citenamefont {Goodman}}]{Foreman_Mackey_2013}%
  \BibitemOpen
  \bibfield  {author} {\bibinfo {author} {\bibfnamefont {D.}~\bibnamefont
  {Foreman-Mackey}}, \bibinfo {author} {\bibfnamefont {D.~W.}\ \bibnamefont
  {Hogg}}, \bibinfo {author} {\bibfnamefont {D.}~\bibnamefont {Lang}},\ and\
  \bibinfo {author} {\bibfnamefont {J.}~\bibnamefont {Goodman}},\ }\bibfield
  {title} {\bibinfo {title} {emcee: The mcmc hammer},\ }\href
  {https://doi.org/10.1086/670067} {\bibfield  {journal} {\bibinfo  {journal}
  {Publications of the Astronomical Society of the Pacific}\ }\textbf {\bibinfo
  {volume} {125}},\ \bibinfo {pages} {306–312} (\bibinfo {year}
  {2013})}\BibitemShut {NoStop}%
\bibitem [{\citenamefont {Alwall}\ \emph {et~al.}(2014)\citenamefont {Alwall},
  \citenamefont {Frederix}, \citenamefont {Frixione}, \citenamefont {Hirschi},
  \citenamefont {Maltoni}, \citenamefont {Mattelaer}, \citenamefont {Shao},
  \citenamefont {Stelzer}, \citenamefont {Torrielli},\ and\ \citenamefont
  {Zaro}}]{Alwall:2014hca}%
  \BibitemOpen
  \bibfield  {author} {\bibinfo {author} {\bibfnamefont {J.}~\bibnamefont
  {Alwall}}, \bibinfo {author} {\bibfnamefont {R.}~\bibnamefont {Frederix}},
  \bibinfo {author} {\bibfnamefont {S.}~\bibnamefont {Frixione}}, \bibinfo
  {author} {\bibfnamefont {V.}~\bibnamefont {Hirschi}}, \bibinfo {author}
  {\bibfnamefont {F.}~\bibnamefont {Maltoni}}, \bibinfo {author} {\bibfnamefont
  {O.}~\bibnamefont {Mattelaer}}, \bibinfo {author} {\bibfnamefont {H.~S.}\
  \bibnamefont {Shao}}, \bibinfo {author} {\bibfnamefont {T.}~\bibnamefont
  {Stelzer}}, \bibinfo {author} {\bibfnamefont {P.}~\bibnamefont {Torrielli}},\
  and\ \bibinfo {author} {\bibfnamefont {M.}~\bibnamefont {Zaro}},\ }\bibfield
  {title} {\bibinfo {title} {{The automated computation of tree-level and
  next-to-leading order differential cross sections, and their matching to
  parton shower simulations}},\ }\href
  {https://doi.org/10.1007/JHEP07(2014)079} {\bibfield  {journal} {\bibinfo
  {journal} {JHEP}\ }\textbf {\bibinfo {volume} {07}},\ \bibinfo {pages}
  {079}},\ \Eprint {https://arxiv.org/abs/1405.0301} {arXiv:1405.0301 [hep-ph]}
  \BibitemShut {NoStop}%
\bibitem [{\citenamefont {Sjöstrand}\ \emph {et~al.}(2015)\citenamefont
  {Sjöstrand}, \citenamefont {Ask}, \citenamefont {Christiansen},
  \citenamefont {Corke}, \citenamefont {Desai}, \citenamefont {Ilten},
  \citenamefont {Mrenna}, \citenamefont {Prestel}, \citenamefont {Rasmussen},\
  and\ \citenamefont {Skands}}]{Sjostrand:2014zea}%
  \BibitemOpen
  \bibfield  {author} {\bibinfo {author} {\bibfnamefont {T.}~\bibnamefont
  {Sjöstrand}}, \bibinfo {author} {\bibfnamefont {S.}~\bibnamefont {Ask}},
  \bibinfo {author} {\bibfnamefont {J.~R.}\ \bibnamefont {Christiansen}},
  \bibinfo {author} {\bibfnamefont {R.}~\bibnamefont {Corke}}, \bibinfo
  {author} {\bibfnamefont {N.}~\bibnamefont {Desai}}, \bibinfo {author}
  {\bibfnamefont {P.}~\bibnamefont {Ilten}}, \bibinfo {author} {\bibfnamefont
  {S.}~\bibnamefont {Mrenna}}, \bibinfo {author} {\bibfnamefont
  {S.}~\bibnamefont {Prestel}}, \bibinfo {author} {\bibfnamefont {C.~O.}\
  \bibnamefont {Rasmussen}},\ and\ \bibinfo {author} {\bibfnamefont {P.~Z.}\
  \bibnamefont {Skands}},\ }\bibfield  {title} {\bibinfo {title} {{An
  Introduction to PYTHIA 8.2}},\ }\href
  {https://doi.org/10.1016/j.cpc.2015.01.024} {\bibfield  {journal} {\bibinfo
  {journal} {Comput. Phys. Commun.}\ }\textbf {\bibinfo {volume} {191}},\
  \bibinfo {pages} {159} (\bibinfo {year} {2015})},\ \Eprint
  {https://arxiv.org/abs/arXiv:1410.3012} {arXiv:arXiv:1410.3012 [hep-ph]}
  \BibitemShut {NoStop}%
\bibitem [{\citenamefont {de~Favereau}\ \emph {et~al.}(2014)\citenamefont
  {de~Favereau}, \citenamefont {Delaere}, \citenamefont {Demin}, \citenamefont
  {Giammanco}, \citenamefont {Lemaître}, \citenamefont {Mertens},\ and\
  \citenamefont {Selvaggi}}]{deFavereau:2013fsa}%
  \BibitemOpen
  \bibfield  {author} {\bibinfo {author} {\bibfnamefont {J.}~\bibnamefont
  {de~Favereau}}, \bibinfo {author} {\bibfnamefont {C.}~\bibnamefont
  {Delaere}}, \bibinfo {author} {\bibfnamefont {P.}~\bibnamefont {Demin}},
  \bibinfo {author} {\bibfnamefont {A.}~\bibnamefont {Giammanco}}, \bibinfo
  {author} {\bibfnamefont {V.}~\bibnamefont {Lemaître}}, \bibinfo {author}
  {\bibfnamefont {A.}~\bibnamefont {Mertens}},\ and\ \bibinfo {author}
  {\bibfnamefont {M.}~\bibnamefont {Selvaggi}} (\bibinfo {collaboration}
  {DELPHES 3}),\ }\bibfield  {title} {\bibinfo {title} {{DELPHES 3, A modular
  framework for fast simulation of a generic collider experiment}},\ }\href
  {https://doi.org/10.1007/JHEP02(2014)057} {\bibfield  {journal} {\bibinfo
  {journal} {JHEP}\ }\textbf {\bibinfo {volume} {02}},\ \bibinfo {pages}
  {057}},\ \Eprint {https://arxiv.org/abs/1307.6346} {arXiv:1307.6346 [hep-ex]}
  \BibitemShut {NoStop}%
\bibitem [{\citenamefont {Goebel}\ \emph {et~al.}(2005)\citenamefont {Goebel},
  \citenamefont {Dawy}, \citenamefont {Hagenauer},\ and\ \citenamefont
  {Mueller}}]{mutual_info_asympt}%
  \BibitemOpen
  \bibfield  {author} {\bibinfo {author} {\bibfnamefont {B.}~\bibnamefont
  {Goebel}}, \bibinfo {author} {\bibfnamefont {Z.}~\bibnamefont {Dawy}},
  \bibinfo {author} {\bibfnamefont {J.}~\bibnamefont {Hagenauer}},\ and\
  \bibinfo {author} {\bibfnamefont {J.}~\bibnamefont {Mueller}},\ }\bibfield
  {title} {\bibinfo {title} {An approximation to the distribution of finite
  sample size mutual information estimates}\ }(\bibinfo {year} {2005})\ pp.\
  \bibinfo {pages} {1102 -- 1106 Vol. 2}\BibitemShut {NoStop}%
\bibitem [{\citenamefont {Baker}\ and\ \citenamefont
  {Cousins}(1984)}]{BAKER1984437}%
  \BibitemOpen
  \bibfield  {author} {\bibinfo {author} {\bibfnamefont {S.}~\bibnamefont
  {Baker}}\ and\ \bibinfo {author} {\bibfnamefont {R.~D.}\ \bibnamefont
  {Cousins}},\ }\bibfield  {title} {\bibinfo {title} {Clarification of the use
  of chi-square and likelihood functions in fits to histograms},\ }\href
  {https://doi.org/https://doi.org/10.1016/0167-5087(84)90016-4} {\bibfield
  {journal} {\bibinfo  {journal} {Nuclear Instruments and Methods in Physics
  Research}\ }\textbf {\bibinfo {volume} {221}},\ \bibinfo {pages} {437}
  (\bibinfo {year} {1984})}\BibitemShut {NoStop}%
\end{thebibliography}%


\end{document}